\documentclass[trackchanges]{aastex701}

\usepackage{overpic}
\usepackage{xcolor} 

\begin{document}

\title{Design, Testing, and Commissioning of the Sun Yat-sen University (SYSU) 80\,cm Infrared Telescope}

\author[orcid=0009-0003-5592-3734]{Zhong-Nan Dong}
\affiliation{School of Physics and Astronomy, Sun Yat-sen University, Zhuhai 519082, China}
\affiliation{CSST Science Center for the Guangdong-Hong Kong-Macau Greater Bay Area, Sun Yat-sen University, Zhuhai 519082, China}
\affiliation{National Astronomical Observatories, Chinese Academy of Sciences, Beijing 100101, China}
\email{dongzhn@mail2.sysu.edu.cn}

\author[orcid=0000-0002-6077-6287]{Bin Ma}
\affiliation{School of Physics and Astronomy, Sun Yat-sen University, Zhuhai 519082, China}
\affiliation{CSST Science Center for the Guangdong-Hong Kong-Macau Greater Bay Area, Sun Yat-sen University, Zhuhai 519082, China}
\email{mabin3@mail.sysu.edu.cn}
\correspondingauthor{Bin Ma}
\email{mabin3@mail.sysu.edu.cn}

\author[orcid=0009-0004-4767-3146]{Chun Chen}
\affiliation{School of Physics and Astronomy, Sun Yat-sen University, Zhuhai 519082, China}
\affiliation{CSST Science Center for the Guangdong-Hong Kong-Macau Greater Bay Area, Sun Yat-sen University, Zhuhai 519082, China}
\affiliation{Dipartimento di Fisica, Universit$\grave{a}$ di Napoli “Federico II”, Compl. Univ. di Monte S. Angelo, Via Cinthia, I-80126, Napoli, Italy}
\email{chench386@mail2.sysu.edu.cn}

\author[orcid=0009-0006-2501-139X]{Wei-Sen Huang}
\affiliation{School of Physics and Astronomy, Sun Yat-sen University, Zhuhai 519082, China}
\affiliation{CSST Science Center for the Guangdong-Hong Kong-Macau Greater Bay Area, Sun Yat-sen University, Zhuhai 519082, China}
\affiliation{Yunnan Observatories, Chinese Academy of Sciences, Kunming 650011, China}
\email{huangws5@mail2.sysu.edu.cn}

\author[orcid=0009-0000-6955-0594]{Jin-Ji Li}
\affiliation{School of Physics and Astronomy, Sun Yat-sen University, Zhuhai 519082, China}
\affiliation{CSST Science Center for the Guangdong-Hong Kong-Macau Greater Bay Area, Sun Yat-sen University, Zhuhai 519082, China}
\email{lijj328@mail2.sysu.edu.cn}

\author[orcid=0000-0002-3134-9526]{Jia-Qi Lin}
\affiliation{School of Physics and Astronomy, Sun Yat-sen University, Zhuhai 519082, China}
\affiliation{CSST Science Center for the Guangdong-Hong Kong-Macau Greater Bay Area, Sun Yat-sen University, Zhuhai 519082, China}
\email{linjq63@mail2.sysu.edu.cn}

\author[orcid=0009-0003-0141-5793]{Yun Shi}
\affiliation{School of Physics and Astronomy, Sun Yat-sen University, Zhuhai 519082, China}
\affiliation{CSST Science Center for the Guangdong-Hong Kong-Macau Greater Bay Area, Sun Yat-sen University, Zhuhai 519082, China}
\email{shiy228@mail2.sysu.edu.cn}

\author[orcid=0000-0003-1412-2028]{Hao-Ran Zhang}
\affiliation{School of Physics and Astronomy, Sun Yat-sen University, Zhuhai 519082, China}
\affiliation{CSST Science Center for the Guangdong-Hong Kong-Macau Greater Bay Area, Sun Yat-sen University, Zhuhai 519082, China}
\email{zhanghr33@mail2.sysu.edu.cn}

\author[orcid=0009-0007-9545-2550]{Duo-Le Cao}
\affiliation{School of Physics and Astronomy, Sun Yat-sen University, Zhuhai 519082, China}
\affiliation{CSST Science Center for the Guangdong-Hong Kong-Macau Greater Bay Area, Sun Yat-sen University, Zhuhai 519082, China}
\email{caodle@mail2.sysu.edu.cn}

\author{Bao-Gang Chen}
\affiliation{Changchun Institute of Optics, Fine Mechanics and Physics, Chinese Academy of Sciences, Changchun 130033, China}
\email{cbg0813@163.com}
\correspondingauthor{Bao-Gang Chen}
\email{cbg0813@163.com}

\author{Tai-Ran Deng}
\affiliation{School of Atmospheric Sciences, Sun Yat-sen University, Zhuhai 519082, China}
\email{dengtr@mail2.sysu.edu.cn}

\author{Rui-Chen Gao}
\affiliation{School of Physics and Astronomy, Sun Yat-sen University, Zhuhai 519082, China}
\affiliation{CSST Science Center for the Guangdong-Hong Kong-Macau Greater Bay Area, Sun Yat-sen University, Zhuhai 519082, China}
\email{gaorch5@mail2.sysu.edu.cn}

\author[orcid=0000-0003-3317-4771]{Yi Hu}
\affiliation{National Astronomical Observatories, Chinese Academy of Sciences, Beijing 100101, China}
\email{huyi.naoc@gmail.com}

\author{Hong-Zhuang Li}
\affiliation{Changchun Institute of Optics, Fine Mechanics and Physics, Chinese Academy of Sciences, Changchun 130033, China}
\email{jilinbayan@163.com}

\author[orcid=0000-0001-6820-1683]{Xia Li}
\affiliation{School of Physics and Astronomy, Sun Yat-sen University, Zhuhai 519082, China}
\affiliation{CSST Science Center for the Guangdong-Hong Kong-Macau Greater Bay Area, Sun Yat-sen University, Zhuhai 519082, China}
\email{lixia76@mail2.sysu.edu.cn}

\author[orcid=0009-0007-2850-9908]{Pu Lin}
\affiliation{School of Physics and Astronomy, Sun Yat-sen University, Zhuhai 519082, China}
\affiliation{CSST Science Center for the Guangdong-Hong Kong-Macau Greater Bay Area, Sun Yat-sen University, Zhuhai 519082, China}
\email{linp@mail2.sysu.edu.cn}

\author{Yang Liu}
\affiliation{Changchun Institute of Optics, Fine Mechanics and Physics, Chinese Academy of Sciences, Changchun 130033, China}
\email{liuyang@ciomp.ac.cn}

\author[orcid=0000-0002-0378-2023]{Bo Ma}
\affiliation{School of Physics and Astronomy, Sun Yat-sen University, Zhuhai 519082, China}
\affiliation{CSST Science Center for the Guangdong-Hong Kong-Macau Greater Bay Area, Sun Yat-sen University, Zhuhai 519082, China}
\email{mabo8@mail.sysu.edu.cn}

\author[orcid=0000-0001-5012-2362]{Rong-Feng Shen}
\affiliation{School of Physics and Astronomy, Sun Yat-sen University, Zhuhai 519082, China}
\affiliation{CSST Science Center for the Guangdong-Hong Kong-Macau Greater Bay Area, Sun Yat-sen University, Zhuhai 519082, China}
\email{shenrf3@mail.sysu.edu.cn}

\author[orcid=0009-0004-3308-3559]{Li-Duo Song}
\affiliation{Changchun Institute of Optics, Fine Mechanics and Physics, Chinese Academy of Sciences, Changchun 130033, China}
\email{songliduo@ciomp.ac.cn}

\author[orcid=0000-0003-2963-4405]{Fang-Yu Xu}
\affiliation{Yunnan Observatories, Chinese Academy of Sciences, Kunming 650011, China}
\email{xu\_fangyu@ynao.ac.cn}
\correspondingauthor{Fang-Yu Xu}
\email{xu\_fangyu@ynao.ac.cn}

\author[orcid=0009-0007-2408-9221]{Hao-Nan Yang}
\affiliation{School of Physics and Astronomy, Sun Yat-sen University, Zhuhai 519082, China}
\affiliation{CSST Science Center for the Guangdong-Hong Kong-Macau Greater Bay Area, Sun Yat-sen University, Zhuhai 519082, China}
\email{yanghn8@mail2.sysu.edu.cn}

\author[orcid=0000-0002-7210-8104]{Yan Yu}
\affiliation{School of Physics and Astronomy, Sun Yat-sen University, Zhuhai 519082, China}
\affiliation{CSST Science Center for the Guangdong-Hong Kong-Macau Greater Bay Area, Sun Yat-sen University, Zhuhai 519082, China}
\affiliation{National Astronomical Observatories, Chinese Academy of Sciences, Beijing 100101, China}
\email{yuyan35@mail2.sysu.edu.cn}

\author{Jun Yuan}
\affiliation{Yunnan Observatories, Chinese Academy of Sciences, Kunming 650011, China}
\email{yuanjun@ynao.ac.cn}

\author[orcid=0000-0002-0551-5615]{Xiang-Tao Zeng}
\affiliation{School of Physics and Astronomy, Sun Yat-sen University, Zhuhai 519082, China}
\affiliation{CSST Science Center for the Guangdong-Hong Kong-Macau Greater Bay Area, Sun Yat-sen University, Zhuhai 519082, China}
\email{zengxt27@mail2.sysu.edu.cn}

\author{Hao-Yuan Zheng}
\affiliation{School of Physics and Astronomy, Sun Yat-sen University, Zhuhai 519082, China}
\affiliation{CSST Science Center for the Guangdong-Hong Kong-Macau Greater Bay Area, Sun Yat-sen University, Zhuhai 519082, China}
\email{zhenghy96@mail2.sysu.edu.cn}

\begin{abstract}

The Sun Yat-sen University (SYSU) 80\,cm telescope is a new generation near-infrared (NIR) facility in China dedicated to time-domain astronomy, while also serving as a testbed for emerging NIR cameras. Commissioned in October 2024 at the 4100\,m Lenghu site on the Tibetan Plateau in China, the telescope adopts a reflective Cassegrain design with two Nasmyth foci for $J$ and $K$ bands. The $J$ band imaging system, initially equipped with a $640 \times 512$ off-the-shelf InGaAs camera (INS Mars640) and upgraded in June 2025 to a $1280 \times 1024$ science-grade, deeply cooled camera (YNAOIR), achieves background-limited performance with a dark current of $\sim$\,14\,e$^{-}$\,s$^{-1}$\,pix$^{-1}$ and a readout noise of $\sim$\,11\,e$^{-}$. The system reaches a limiting magnitude of $J \sim 17$\,mag (Vega system) in single 20\,s exposures and depths of $J \sim 19.4$\,mag with stacked 30\,minute exposures. For a variable with $J \sim 14$\,mag during on-sky tests, the system delivers millimagnitude-level photometric precision. Since commissioning, the telescope observed transients such as gamma-ray bursts (GRBs), supernovae and comets, variables including active galactic nuclei (AGNs), high-redshift quasars ($z \gtrsim 6$), and brown dwarfs, as well as deep-field imaging reaching $J \sim 20.5$\,mag. This validates the feasibility of using InGaAs cameras for astronomical observations, encouraging other institutions to develop dedicated infrared telescopes or integrate infrared cameras into existing optical telescopes.

\end{abstract}


\section{Introduction} \label{sec:Intro}

Near-infrared (NIR) time-domain observations are essential for exploring the dynamic universe, specifically to overcome fundamental limitations of optical surveys regarding cosmological redshift, dust extinction, and intrinsic temperatures.
First, cosmological redshift mandates NIR observations to probe the early universe. For high-redshift objects ($z \gtrsim 6$), the presence of the Lyman break results in a sharp drop in flux at optical wavelengths. For high-redshift Gamma-ray bursts (GRBs), this makes NIR identification the primary method for recognizing these sources \citep{2009Natur.461.1258S}. For high-redshift quasars, NIR photometry is indispensable for sampling their intrinsic spectral energy distributions and confirming photometric redshifts \citep{2011Natur.474..616M, 2015Natur.518..512W}.
Second, by mitigating dust extinction, NIR observations enable precision cosmology using Type Ia supernovae (SNe Ia). SNe Ia exhibit higher luminosity uniformity and reduced intrinsic scatter in the NIR \citep{2006ApJ...649..939K}. These characteristics minimize systematic errors in the Hubble diagram, resulting in tighter constraints on dark energy \citep{2016NatSR...635596N}.
Finally, the NIR regime captures the peak flux of the coolest stellar and substellar objects. For ultracool dwarfs (late-M and later, $\le$\,3000\,K), high-precision NIR photometry reveals critical atmospheric molecular features, such as H$_2$O, CH$_4$ \citep{2004AJ....127.3553K}. M-type dwarfs are also prime targets for detecting terrestrial planets within their habitable zones due to their high abundance and longevity \citep{2016Natur.533..221G, 2024ApJ...976...82T}.

Despite the growing scientific interest in the dynamic infrared sky, dedicated ground-based near-infrared time-domain facilities remain rare. While many observatories are equipped with NIR instruments, most are general-purpose facilities that allocate only limited time to continuous monitoring. Among the few specialized systems, the Simultaneous 3-color InfraRed Imager for Unbiased Survey (SIRIUS; \citealt{2003SPIE.4841..459N}) and its successor, Kyoto-SIRIUS (kSIRIUS; \citealt{2024SPIE13096E..3IN}), provide simultaneous multi-band monitoring over small fields of view, optimized for targeted high-cadence observations. For wide-field operations, the Palomar Gattini-IR (PGIR; \citealt{2020PASP..132b5001D}), the Wide-Field Infrared Transient Explorer (WINTER; \citealt{2025arXiv251216753F}), and the PRime-focus Infrared Microlensing Experiment (PRIME; \citealt{2025AJ....170..338S}) have recently joined the field, dedicated primarily to large-scale time-domain surveys. Complementing these dedicated facilities, general-purpose telescopes play a critical role in transient follow-up. Large-aperture observatories, such as Gemini North (NIRI; \citealt{2003PASP..115.1388H}) and the Hale Telescope (WIRC; \citealt{2003SPIE.4841..451W}), provide essential depth, reaching limiting magnitudes of $J \sim 23.0$--$23.5$\,mag (Vega system, hereafter; 5$\sigma$, 1\,hr) \citep{2003PASP..115.1388H, 2003SPIE.4841..451W}. Medium-class (2--4\,m) facilities, including the Visible and Infrared Survey Telescope for Astronomy (VISTA; \citealt{2015A&A...575A..25S}), the Telescopio Nazionale Galileo (TNG; \citealt{1997SPIE.2871..244B}), the Nordic Optical Telescope (NOT; \citealt{2010ASSP...14..211D}), and the Wendelstein 2.1\,m Telescope (3KK; \citealt{2016SPIE.9908E..44L}), contribute vital photometric and spectroscopic capabilities. Additionally, the Catching Optical and Infrared Bright transients for identification and Redshift determination (COLIBRI) telescope, part of the Space-based multi-band astronomical Variable Objects Monitor (SVOM) ground segment, specializes in ultra-rapid follow-up with a slew time of under 20\,s to any target \citep{2022SPIE12182E..1SB, 2023ExA....56..645N}.

Mercury cadmium telluride (HgCdTe) detectors have been the dominant choice in infrared astronomy for more than two decades. As shown in Table~\ref{tab:cameras}, their low readout noise, low dark current, large-format arrays, and tunable bandgap (1.0--4.0\,$\mu$m) make them the preferred choice for infrared telescopes. However, their high manufacturing cost limits the number of infrared survey instruments currently in operation, particularly for small telescopes below the one-meter class. In recent years, with the advancement of indium gallium arsenide (InGaAs) detectors, a more affordable and operationally convenient alternative has emerged for the 0.9--1.7\,$\mu$m wavelength range (corresponding to $J$ band and part of $H$ band). 
While InGaAs detectors typically exhibit higher readout noise and dark current than HgCdTe, their performance is sufficient to remain below the near-infrared sky background level (typically $\sim$16\,mag\,arcsec$^{-2}$ at premier sites, corresponding to tens to hundreds of electrons per pixel; see Section~\ref{sec:camera}). Moreover, the emergence of deeply cooled InGaAs cameras has further narrowed the performance gap relative to HgCdTe detectors, as shown in Table~\ref{tab:cameras}. Therefore, astronomers have begun to use InGaAs detectors to carry out observations in NIR. For example, WINTER employs six 1920\,$\times$\,1080 InGaAs cameras on a 1\,m telescope with a 1\,deg$^{2}$ field of view (FoV), operating at the Palomar Observatory in the Northern Hemisphere since June 2023 \citep{2025arXiv251216753F}. The Antarctic Infrared Binocular Telescope (AIRBT) utilizes dual 640\,$\times$\,512 InGaAs cameras mounted on a 15\,cm binocular at Dome A, Antarctica \citep{2025PASP..137l5003D}. It has conducted $J$ and $H$ band surveys since 2023, and following a January 2025 upgrade, incorporated a specialized $W'$ band filter (1.4\,$\mu$m) to exploit the unique Antarctic atmospheric window \citep{2026arXiv260113068L}.
The Dynamic REd All-sky Monitoring Survey \citep[DREAMS;][]{2020SPIE11203E..07S, 2022JATIS...8a6001B} uses six 1280\,$\times$\,1024 InGaAs cameras coupled with a 0.5\,m telescope at Siding Spring Observatory in the Southern Hemisphere, with commissioning planned for 2026.

In China, there had been no dedicated NIR astronomical telescopes for decades, since the 1.26 m infrared telescope \citep{1988PBeiO..11...33Q} was repurposed for optical observations. In response to this long-standing scarcity and the rapid advancement of InGaAs detectors, we have built and operate the Sun Yat-sen University (SYSU) 80\,cm near-infrared telescope. The telescope uses an InGaAs detector for $J$ band, which also serves as a technological testbed for emerging infrared detector technologies. An HgCdTe detector is installed for $K$ band. The telescope is located at Lenghu on the Tibetan Plateau, a site providing the conditions for near-infrared observations \citep{2021Natur.596..353D}, and entered its commissioning phase in October 2024.

Our primary scientific goals focus on infrared time-domain astronomy: the study of transients such as gamma-ray burst afterglows, supernovae, and comets; the study of variable objects including active galactic nuclei (AGNs) and brown dwarfs; and static deep-field imaging for targets of interest. Meanwhile, it serves as a technological testbed for the astronomical application of InGaAs cameras, providing scientific recommendations for the development of new systems.
This is the new generation ground-based telescope in China equipped with imaging capabilities for the near-infrared $J$ and $K$ bands.
This paper describes the telescope’s system architecture, on-sky performance, and preliminary scientific results. Section \ref{sec:system} outlines the telescope design, the laboratory characterization of the $J$ band InGaAs cameras, as well as the observatory site and meteorological support system. Section \ref{sec:on-sky} presents the on-sky imaging quality, astrometric performance, limiting magnitudes, photometric accuracy, noise analysis, and system stability. Section \ref{sec:scientific} introduces the preliminary scientific observations.

\section{System Overview}\label{sec:system}

\begin{figure*}[ht]
\centering

% 第一行：原图1的内容 (a, b)
\begin{overpic}[scale=0.125]{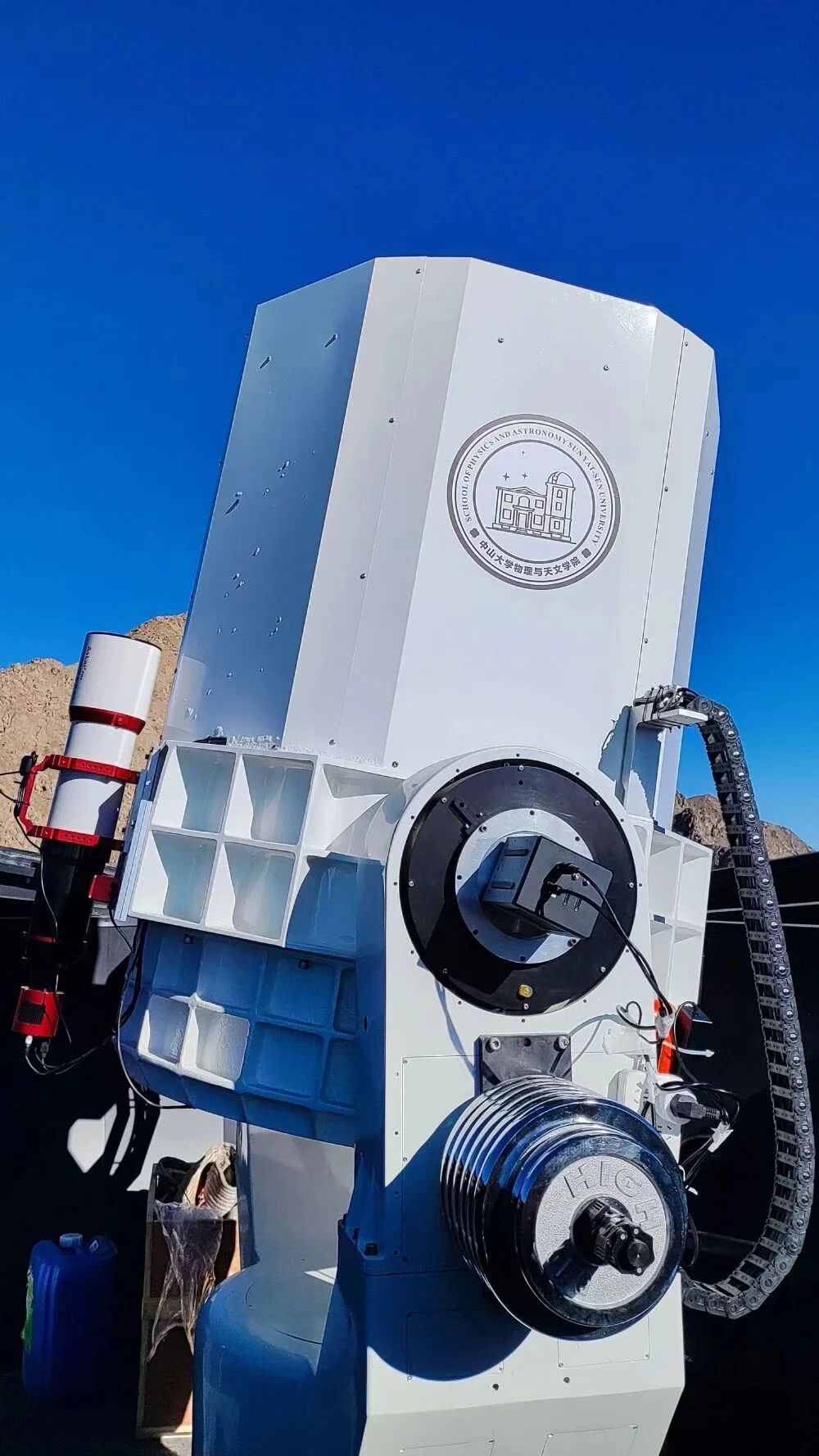}
    \put(1,95){\large \textbf{\color{black}(a)}}
\end{overpic}
\begin{overpic}[scale=0.86]{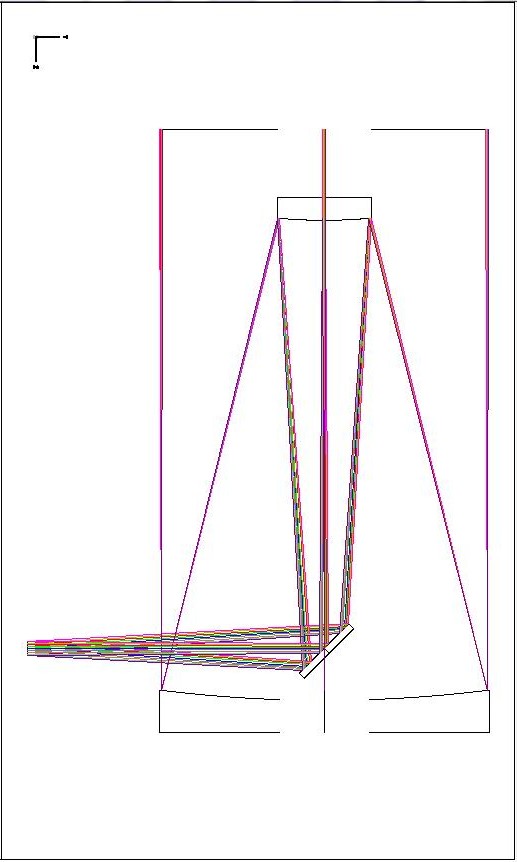}
    \put(51,95){\large \textbf{\color{black}(b)}}
\end{overpic}
\begin{overpic}[scale=0.7]{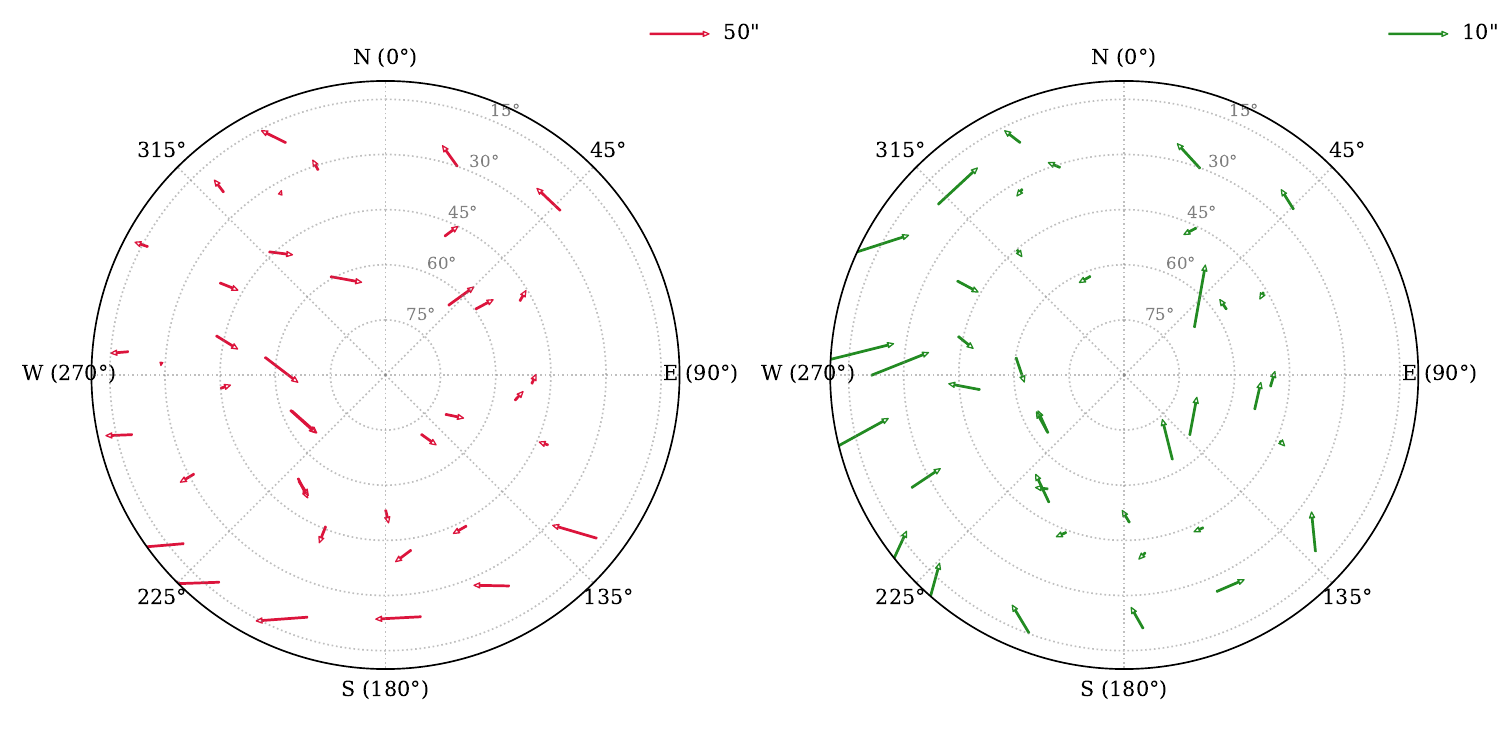}
\put(6,6){\large \textbf{(c)}}
\end{overpic}
\begin{overpic}[scale=0.7]{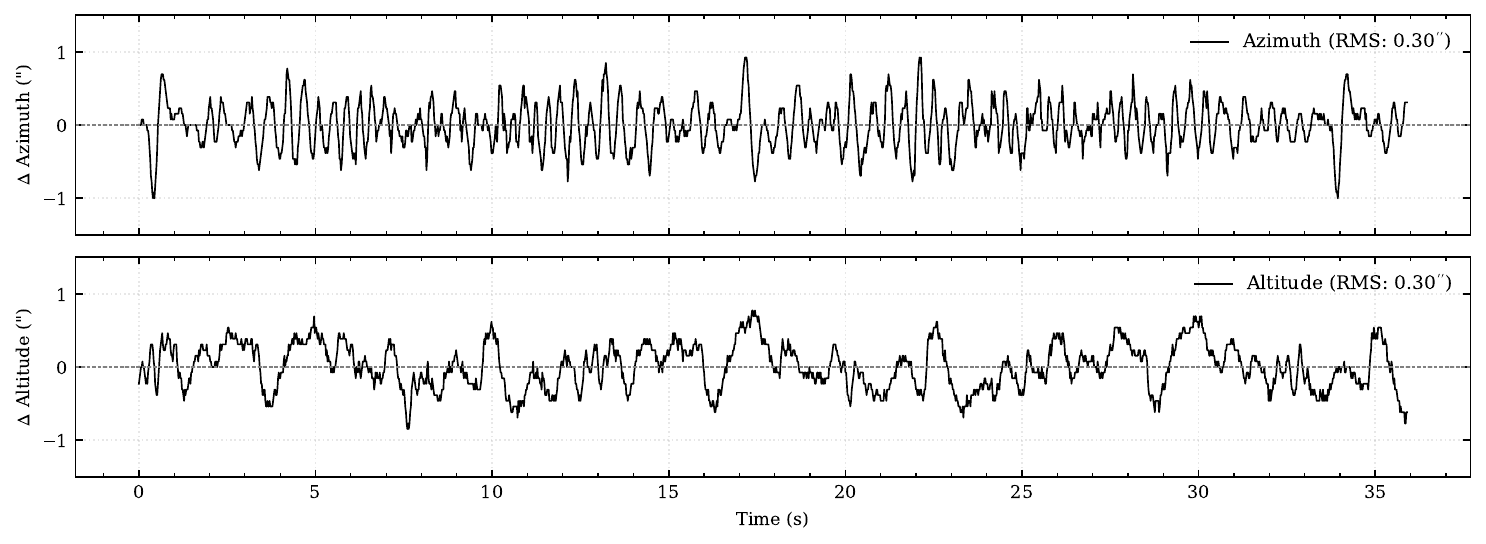}
\put(6,22){\large \textbf{(d)}}
\put(6,6){\large \textbf{(e)}}
\end{overpic}

\caption{
\textbf{(a)} The Sun Yat-sen University 80\,cm infrared telescope at Lenghu on the Tibetan Plateau.
\textbf{(b)} Optical path of the $J$ band system.
\textbf{(c)} Pointing error maps across different azimuth and elevation angles. The left panel shows the uncorrected results (RMS $\sim$ 22.4\arcsec), and the right panel shows the results after pointing model correction (RMS $\sim$ 5.7\arcsec).
\textbf{(d)} and \textbf{(e)} represent the telescope's open-loop tracking accuracy in the azimuth and altitude directions, respectively.
}
\label{fig:telescope_performance}
\end{figure*}

As shown in Fig.~\ref{fig:telescope_performance}~(a), our telescope was installed at Lenghu in October 2024. The optical design of the telescope is optimized for the near-infrared observing conditions at Lenghu, as well as the need for rapid follow-up of transient events. The system supports two instruments operating in $J$ and $K$ bands, using InGaAs and HgCdTe detectors, respectively. The $J$ band camera meets the requirements for high-precision astronomical observations \citep{2024SPIE13103E..1MD, 2025arXiv251014839D}, while the $K$ band camera with scientific grade detectors has been undergoing optimization after the tests with engineering-grade detectors~\citep{Qu2025AIPAdvances}.
Consequently, this paper presents only the results of the $J$ band system, while those for the $K$ band will be reported in future publications after the camera development is completed.

\subsection{Telescope}\label{sec:tel}

\begin{table}[ht]
\centering
\caption{The mirror parameters of the SYSU 80\,cm infrared Telescope}
\label{tab:mirror_params}
\begin{tabular}{lcccc}
\hline
\hline
Mirror & Physical Diameter & Clear Aperture & Radius of Curvature & Conic Constant \\
       & (mm)              & (mm)           & (mm)                &            \\
\hline
M1 (Primary)   & 820  & 800  & 5650 & $-0.910757$ \\
M2 (Secondary) & 240  & 230  & 400  & $-1.210757$ \\
M3 (Tertiary)  & ---  & $230 \times 162$ & Flat & --- \\
\hline
\end{tabular}
\end{table}

The SYSU 80\,cm Infrared Telescope, designed and manufactured by Changchun Institute of Optics, Fine Mechanics and Physics, Chinese Academy of Sciences, employs a focal ratio of $f/8$. This relatively slow focal ratio is optimized for our focus on individual target observations, prioritizing deeper detection and higher spatial sampling over the large field of view. By reducing the contribution from the sky brightness and thermal emission — which is hundreds of times brighter in infrared than in optical — this configuration leads to a lower background noise level, improving the signal-to-noise ratio (S/N) and the detection limits for faint stars. 
Complementing this slow focal ratio, the optical system utilizes a quasi-Ritchey-Chr\'etien (RC) design (Table~\ref{tab:mirror_params}) to provide sharp stellar profiles essential for faint target observations. The secondary mirror assembly introduces a central linear obstruction of $<33\%$. Additionally, all mirrors are made of Zerodur-type glass-ceramic to maintain image quality stability against ambient temperature fluctuations.
The optical design offers a 35\,mm diameter FoV, leaving room for future upgrades to larger-format detectors. The filter set follows the Mauna Kea Observatories near-infrared standards, which are optimized for mid-latitude sites to minimize background emission \citep{2002PASP..114..180T, 2002PASP..114..169S}.
Fig.~\ref{fig:telescope_performance}~(b) illustrates the $J$ band optical path. The telescope adopts a reflective Cassegrain design with Nasmyth foci. After reflection from the primary and secondary mirrors, the beam is directed by a rotatable tertiary mirror to the Nasmyth focus. Focusing is achieved through motorized control of the secondary mirror. Besides, a camera is mounted on the Nasmyth mount, together with the correcting lens and filters. This arrangement allows convenient installation and removal of the cameras.

The telescope is designed with an alt-azimuth mount. Despite the blind region being confined to within roughly $5^\circ$ of the zenith, the system offers flexible pointing and tracking capabilities. During on-sky tests, the telescope achieved a maximum slewing speed of 15$^\circ$\,s$^{-1}$ in azimuth and 6$^\circ$\,s$^{-1}$ in elevation, enabling target acquisition within about 15\,s.

We evaluated the open-loop pointing accuracy of the telescope, as shown in Figure~\ref{fig:telescope_performance}(c). Prior to correction, the Root Mean Square (RMS) pointing error was 22.4\arcsec\ , with maximum deviations reaching up to 45\arcsec\ . To minimize these errors, we applied a 10-term pointing model \citep{1994StaUN.100.....W} defined by:
\begin{eqnarray}
\Delta Az &=& g + b \cdot \tan Alt + c \cdot \sec Alt + IAX \cdot \sin Az \cdot \tan Alt - IAY \cdot \cos Az \cdot \tan Alt \\
\Delta Alt &=& h + IEX \cdot \cos Az + IEY \sin Az + d \cdot \cos Alt + r \cdot \tan Alt
\end{eqnarray}
where $\Delta Az$ and $\Delta Alt$ represent the pointing errors in the azimuth and altitude axes, respectively. The fitted parameters (in arcseconds) are as follows: azimuth zero offset $g = -135.0$, altitude zero offset $h = 4.0$, non-perpendicularity $b = 18.0$, collimation error $c = 90.0$, azimuth-axis tilts $IAX = -15.0$ and $IAY = -10.0$, altitude-axis tilts $IEX = -9.0$ and $IEY = -9.0$, tube flexure $d = -10.0$, and refraction residue $r = 5.0$.
The parameters $g$ and $c$ dominate the error budget. The large offset $g$ originates from the initial mechanical alignment process and does not reflect the inherent optical quality of the telescope. The collimation error $c$ is likely attributed to a slight tilt of the secondary mirror or minor camera mounting inaccuracies. The right panel of Figure~\ref{fig:telescope_performance}(c) displays the residuals after the pointing model correction. The RMS error converges significantly to 5.7\arcsec\ , which is sufficient to reliably place the target near the center of the field of view during open-loop pointing. Notably, the RMS error is less than 2\arcsec\ at elevation angles around $45^\circ$, but the residuals exhibit evidence of tube droop at extremely high and low elevations. This deviation from a simple cosine law for tube flexure is likely due to the heavy, complex cryogenic equipment attached to the Nasmyth mount on the $K$ band side, which introduces asymmetric mechanical deformation to the telescope tube.

The open-loop tracking performance of the telescope is presented in Figures~\ref{fig:telescope_performance}~(d) and (e). The tracking RMS errors in both azimuth and altitude are approximately 0.3\arcsec\ ($< 1$\,pixel). This precision ensures the acquisition of stable images during typical single-frame exposures (20\,s, as detailed in Section~\ref{sec:raytron}).

We evaluated the long-term field derotation accuracy. During a 4.8-hour continuous observation of CSS~J040017.9+382426 on October 1, 2025, the residual field rotation error was measured to be approximately $0.15^\circ$. For a target located at the edge of the $1280 \times 1024$ detector array, this rotational error induced a displacement of 2\,pixels ($\sim 1\arcsec$). Such stability indicates that the derotation accuracy is well within acceptable margins for our routine scientific observations.

\subsection{$J$ Band cameras}\label{sec:camera}

\begin{deluxetable*}{lcccccccc}
\tablecaption{Comparison of Characterization Results\label{tab:cameras}}
\tablehead{
    \colhead{Detector} &
    \colhead{Size} &
    \colhead{Pitch} &
    \colhead{$T_{\rm op}$} &
    \colhead{RN} &
    \colhead{$I_{\rm dark}$} &
    \colhead{FW} &
    \colhead{NL} &
    \colhead{Ref.} \\
    \colhead{} &
    \colhead{(pixels)} &
    \colhead{($\mu$m)} &
    \colhead{(K)} &
    \colhead{(e$^-$)} &
    \colhead{(e$^{-}$\,s$^{-1}$\,pix$^{-1}$)} &
    \colhead{(ke$^-$)} &
    \colhead{(\%)} &
    \colhead{}
}
\startdata
INS Mars640 (HG)\tablenotemark{a} & $640 \times 512$   & 15 & 223 & 14 & \nodata & 16 & $<1$ & This Work \\
INS Mars640 (MG)\tablenotemark{a} & $640 \times 512$   & 15 & 223 & 33 & 221 & 92 & $<1$ & This Work \\
DREAMS                        & $1280 \times 1024$ & 12 & 213 & 90 & $<67$ & 57 & 4 & \cite{2022JATIS...8a6001B} \\
YNAOIR (HG)\tablenotemark{b}  & $1280 \times 1024$ & 15 & 80  & 11 & 14 & 33 & $<1$ & This Work \\
YNAOIR (LG)\tablenotemark{b}  & $1280 \times 1024$ & 15 & 80  & 32 & 44 & 131 & $<1$ & This Work \\
COUGAR                        & $640 \times 512$   & 20 & 77  & $<5$ & $<1$ & $\ge$400 & \nodata & \cite{2014SPIE.9070E..0BV} \\
HxRG                          & $4096 \times 4096$& 1.7--16 & $80$ & 8--20 & 0.01 & 80--150 & 10--25 & \cite{2011ASPC..437..383B} \\
\enddata
\begin{flushleft}
\tablecomments{
Abbreviations: $T_{\rm op}$ = Operating Temperature; RN = Readout Noise; $I_{\rm dark}$ = Dark Current; FW = Full-Well capacity; NL = Non-Linearity.
}
\tablenotetext{a}{High Gain (HG): 1.0\,e$^-$/ADU, Middle Gain (MG): 5.6\,e$^-$/ADU.}
\tablenotetext{b}{High Gain (HG): 0.5\,e$^-$/ADU, Low Gain (LG): 2.0\,e$^-$/ADU.}
\end{flushleft}
\end{deluxetable*}

The system serves as a testing platform for short-wave infrared cameras, allowing us to evaluate various detectors for $J$ band observations. When the commissioning phase began in October 2024, we used a commercial $640 \times 512$ InGaAs camera (effective FoV of 5\arcmin $\times$ 4\arcmin) manufactured by SPLG ShenZhen Technology Co., Ltd. (hereafter the INS Mars640 camera)\footnote{Product information available at: \url{http://www.bnn-splg.cn/product_276062.html}.}. Following detector upgrades, we replaced it in June 2025 with a science-grade $1248 \times 1024$ camera (effective FoV of 10\arcmin $\times$ 8\arcmin) developed by Yunnan Observatories, Chinese Academy of Sciences (hereafter the \textit{YNAOIR} camera). In November 2025, this camera was further upgraded based on its operational performance. We have carried out detailed laboratory tests for both cameras — including measurements of bias level, readout noise, dark current, nonlinearity, the photon transfer curve (PTC), and gain — with the results summarized in Table~\ref{tab:cameras}.

\begin{figure*}
\centering

\begin{overpic}[scale=0.52]{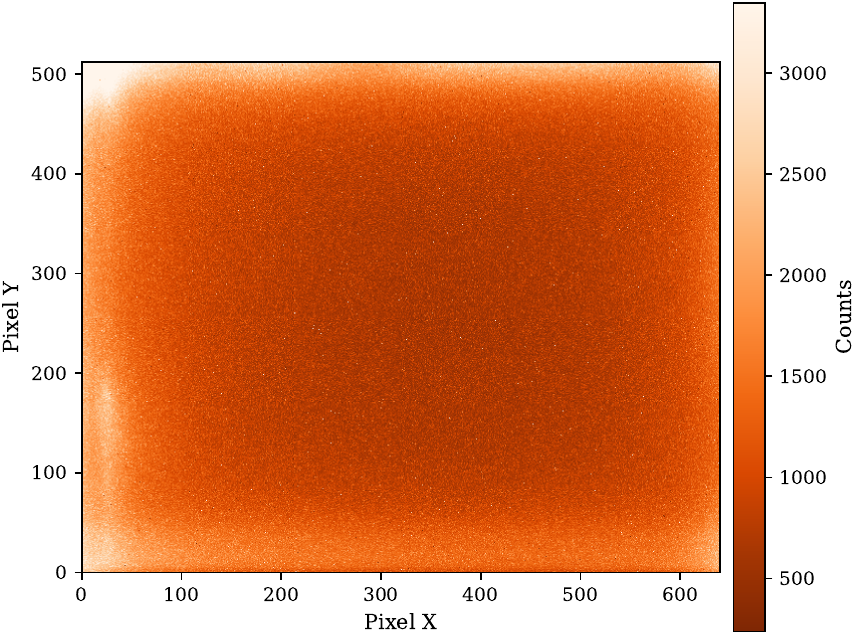}
\end{overpic}
\hspace{-0.175cm}
\begin{overpic}[scale=0.375]{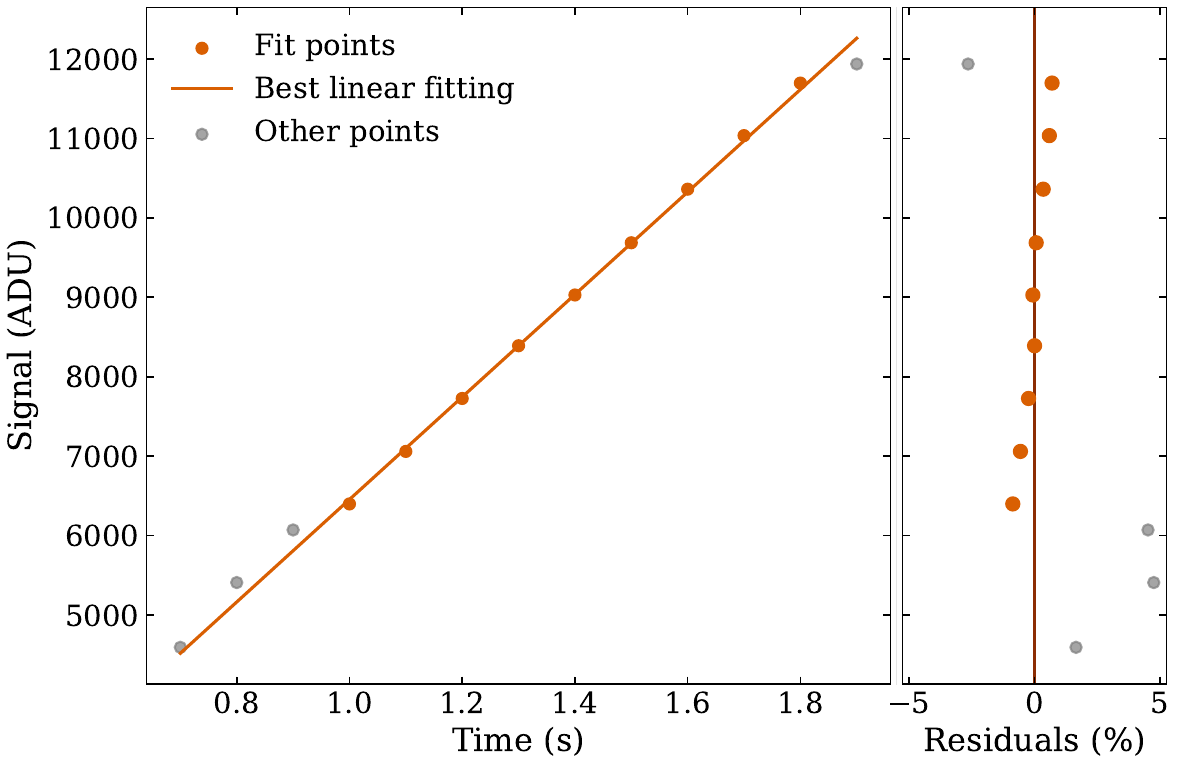}
\end{overpic}

\begin{overpic}[scale=0.525]{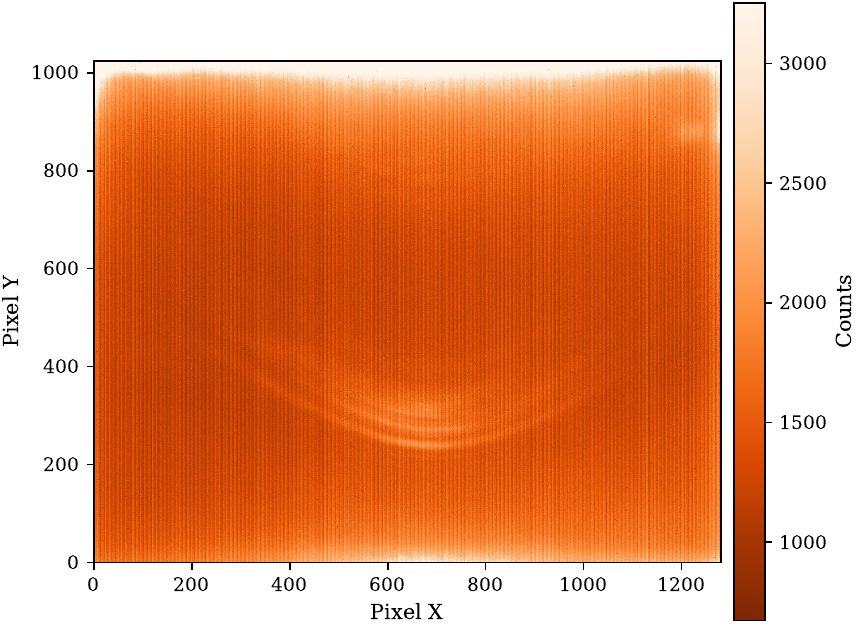}
\end{overpic}
\begin{overpic}[scale=0.375]{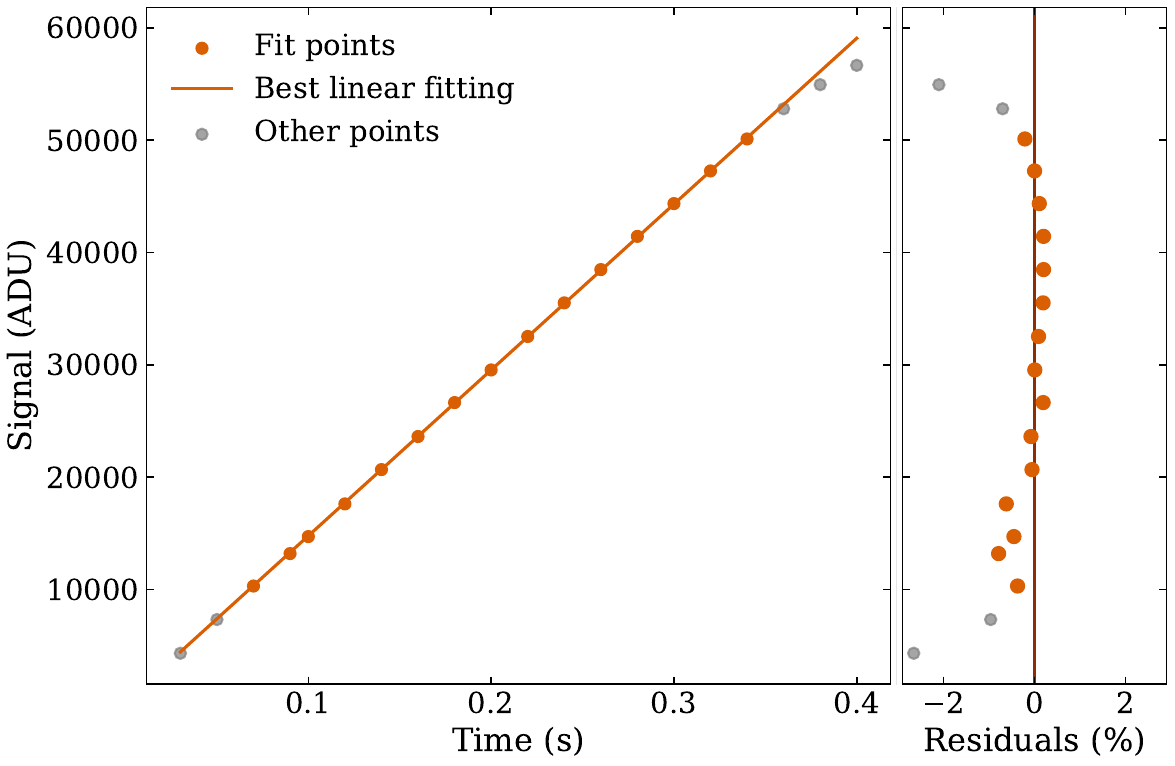}
\end{overpic}

\caption{
Laboratory performance characterization for the INS Mars640 (top row) and YNAOIR (bottom row) cameras, featuring frame noise distributions (left column) and non-linearity tests (right column).
\textbf{For the INS Mars640:} the 20\,s dark current frame (middle gain) shows background noise roughly twice higher in outer $\sim$40-pixel border and up to $\sim$5$\times$ higher in the upper-left corner. Its non-linearity remains below 1\% between 6500 and 11,500\,ADU but rises rapidly beyond this range, reaching $\sim$5\%. 
\textbf{For the YNAOIR:} the 20\,s dark frame (high gain) reveals vertical banding and dark-current-related arc and edge structures. The non-linearity curve shows less than 1\% deviation in the range of 10,000--50,000\,ADU.
}
\label{fig:lab}
\end{figure*}

\subsubsection{The INS Mars640 camera}\label{sec:raytron}

The INS Mars640 camera is a 14-bit, $640 \times 512$ InGaAs system with 15\,$\mu$m pixels. It is thermoelectrically cooled and can maintain an operating temperature of $-50^\circ$C. The camera provides high and middle gain modes, as summarized in Table~\ref{tab:cameras}. In the middle gain mode, it features a readout noise of 33\,$e^{-}$ and a median dark current of 221\,e$^{-}$\,s$^{-1}$\,pix$^{-1}$, which represents a competitive performance level for a commercial-grade InGaAs detector. To balance full-well (FW) capacity (clipped by the digital limit), dynamic range, and operational efficiency, we bypassed the high gain mode in favor of the middle gain mode with a 20\,s exposure time during the commissioning phase. While this configuration is a practical choice for our current observations, we note that a broad range of settings near these values would likely yield comparable scientific results.

The upper left panel of Fig.~\ref{fig:lab} shows a 20\,s dark current exposure obtained in the INS Mars640 middle gain mode. The central region of the detector displays no significant fixed patterns or anomalous structures. Compared to the central region, count levels in the outer $\sim$40-pixel border are 2--5 times higher, thereby introducing correspondingly larger Poisson noise. These spatially coherent regions of elevated signal may originate from electronic glow.
The detector non-linearity remains below 1\% for signal levels between 6500 and 11,500\,ADU. Pixels outside this range require careful calibration because the non-linearity rises rapidly beyond these limits.

\subsubsection{The YNAOIR camera}\label{sec:ynaoir}

The YNAOIR camera is a 16-bit, $1248 \times 1024$ InGaAs system with 15\,$\mu$m pixels. 
The camera incorporates a Stirling cooler within the housing that directly cools the sensor. This design retains the operational simplicity of thermoelectric cooling (TEC) based systems while enabling an operating temperature of 80\,K. The deep-cooled design of the detector suppresses dark current to a level of 14\,e$^{-}$\,s$^{-1}$\,pix$^{-1}$ in the high gain mode. The system provides both high and low gain modes. We find that stacking four high-gain frames (each with exposure time $t$) produces a dynamic range comparable to that of a single low gain frame with exposure time $4t$, while the effective readout noise and dark current Poisson noise are reduced to approximately 0.7 and 0.6 of those in the low gain mode, respectively. We therefore adopt the high gain mode as our default configuration. For consistency and to facilitate direct comparison with the INS Mars640 camera, we use a fixed exposure time of 20\,s throughout.

As illustrated in the lower left panel of Fig.~\ref{fig:lab}, our tests show that the Y-direction stripes structure varies by $\sim$200\,ADU. The arc-shaped bright region near the image center, as well as the enhanced structures in the upper $\sim$100 rows and lower $\sim$30 rows, are brighter than their surroundings by up to $\sim$3000\,ADU. For data reduction, we note that the laboratory-measured bias pattern exhibits residuals of a few percent compared with the on-sky bias structure. To achieve a higher precision characterization of the fixed-pattern background during on-sky observations, we construct a background frame by combining multiple nighttime sky exposures after subtracting their median sky levels. This issue is currently being addressed through ongoing hardware upgrades. The non-linearity of the YNAOIR camera remains below 1\% for signal levels between 10,000 and 50,000\,ADU. Ongoing upgrades will further enhance linearity and broaden the operational range of the linear response.

\subsubsection{Performance Comparison with Similar Systems}\label{sec:cameras_compare}

A comparison of these cameras with other representative near-infrared instruments used in similar programs is presented in Table~\ref{tab:cameras}. While InGaAs detectors generally lag behind HgCdTe arrays in performance—particularly regarding dark current—this limitation is mitigated in ground-based settings. Specifically, the typical $J$ band sky background at mid-latitude observatories is on the order of tens or hundreds of electrons per second per pixel, which often exceeds or is comparable to the detector dark current. Since the sky background typically dominates the total noise budget, InGaAs detectors provide a cost-effective solution without becoming the primary sensitivity bottleneck for most ground-based broad-band imaging applications.
A performance gradient exists even among InGaAs cameras, primarily dictated by cooling. Thermoelectrically cooled systems, such as the INS Mars640 and DREAMS, exhibit dark currents that are close to the sky background, making them suitable for bright targets or high-cadence follow-up of transients in their earliest phases. In contrast, deeply cooled systems like YNAOIR and COUGAR achieve dark currents roughly an order of magnitude lower, shifting the observation into a true sky-limited regime.
Among the deeply cooled cameras, a trade-off exists between performance and practicality. The YNAOIR camera, despite having higher dark current and readout noise than the liquid-nitrogen-cooled COUGAR, offers a larger detector format and the operational convenience of mechanical cooling. This combination makes YNAOIR particularly well-suited for our observational requirements. Its larger FoV facilitates more effective differential photometry by providing a higher density of reference stars—leading to improved relative precision—and enables more extensive spatial coverage for deep-field imaging.

\subsection{The Observatory Site and Meteorological Support System}\label{sec:site}

The Lenghu site offers favorable conditions for astronomical observations in near-infrared. First, the site offers a median seeing of 0.75\arcsec\ in optical, which is likely even better in infrared given the wavelength dependence of atmospheric seeing ($\theta \propto \lambda^{-1/5}$; \citealt{1978JOSA...68..877B}). 
Second, the $J$ band sky brightness was measured to be 15.7\,mag~arcsec$^{-2}$ over a period of more than a month \citep{2024SPIE13094E..5CL}, corresponding to about 65\,e$^{-}$\,s$^{-1}$\,pix$^{-1}$ in our system.

An all-sky camera and a meteorological station — measuring wind speed and direction, temperature, and humidity — have been installed on the observing platform to provide auxiliary support for telescope operations, which are derived from the Kunlun All-sky Camera (KLCAM) and the second-generation Kunlun Automated Weather Station (KLAWS-2G) frameworks \citep{10.1088/1538-3873/aae916,10.1093/mnras/staa3824}, respectively. These systems record, process, and display environmental data in real time, serving as an essential safety safeguard for our unattended operations and laying the groundwork for future fully automated observing.

\section{On-sky performance}\label{sec:on-sky}

To evaluate the observational capabilities of the system, we conducted on-sky tests using two cameras successively. From October 2024 to June 2025, the telescope was equipped with the INS Mars640 camera, after which the system was upgraded to the YNAOIR camera. Our tests included evaluations of image quality, astrometric accuracy, zero point, system efficiency, detection depth and noise characterization, and photometric stability. The typical exposure time for a single frame is 20\,s, as described in Section~\ref{sec:camera}.

\begin{figure*}
\centering

\begin{overpic}[scale=0.31]{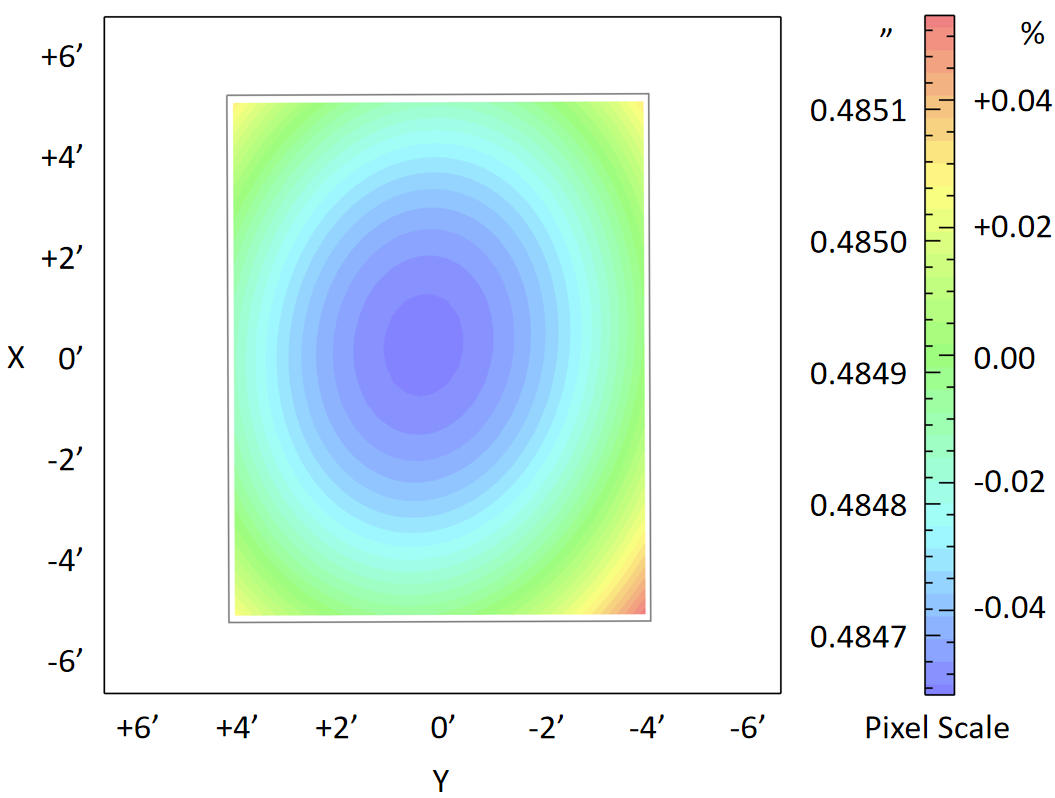}
    \put(11,68){\large \textbf{(a)}}
\end{overpic}
\hspace{0.8cm}
\begin{overpic}[scale=0.2]{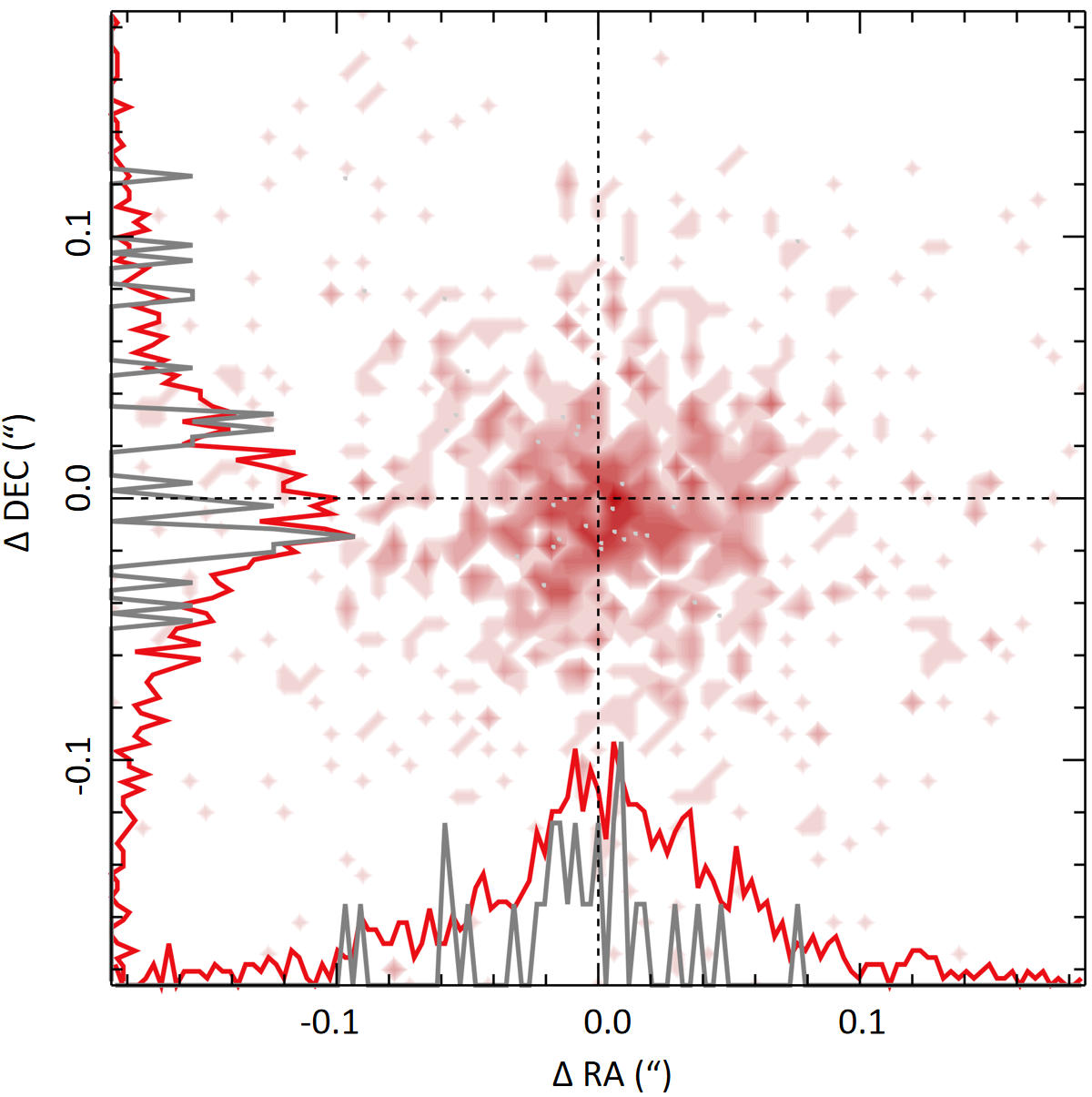}
    \put(87,90.5){\large \textbf{(b)}}
\end{overpic}

\hspace*{0.7cm}
\begin{overpic}[scale=0.485]{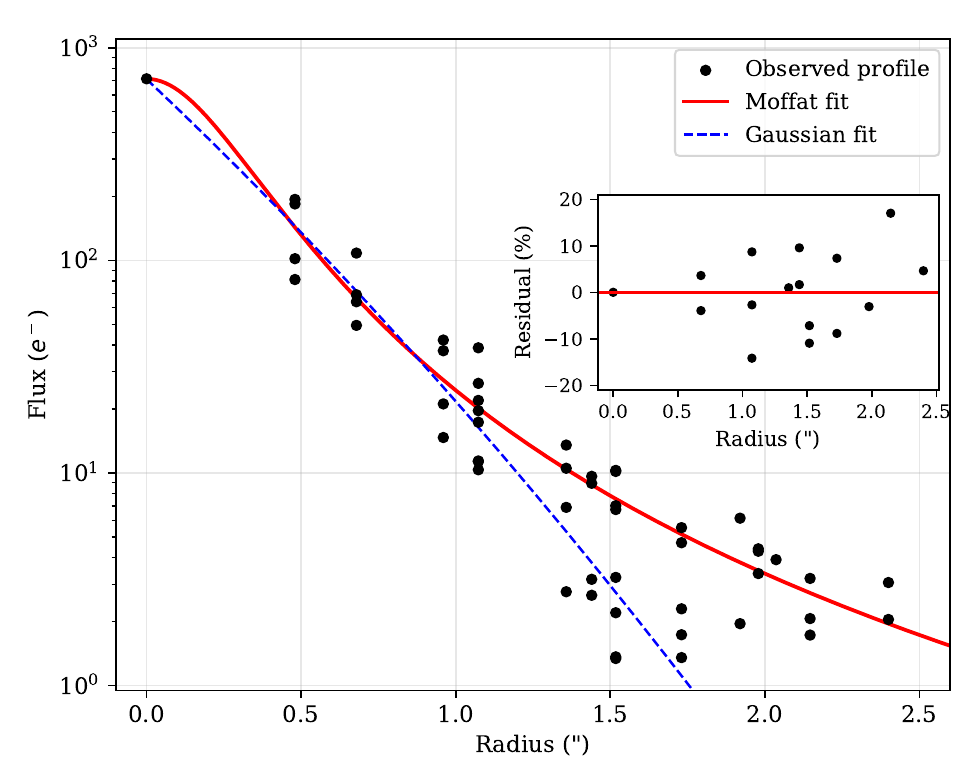}
    \put(13,12.5){\large \textbf{(c)}}
\end{overpic}
\hspace{0.7cm}
\begin{overpic}[scale=0.48]{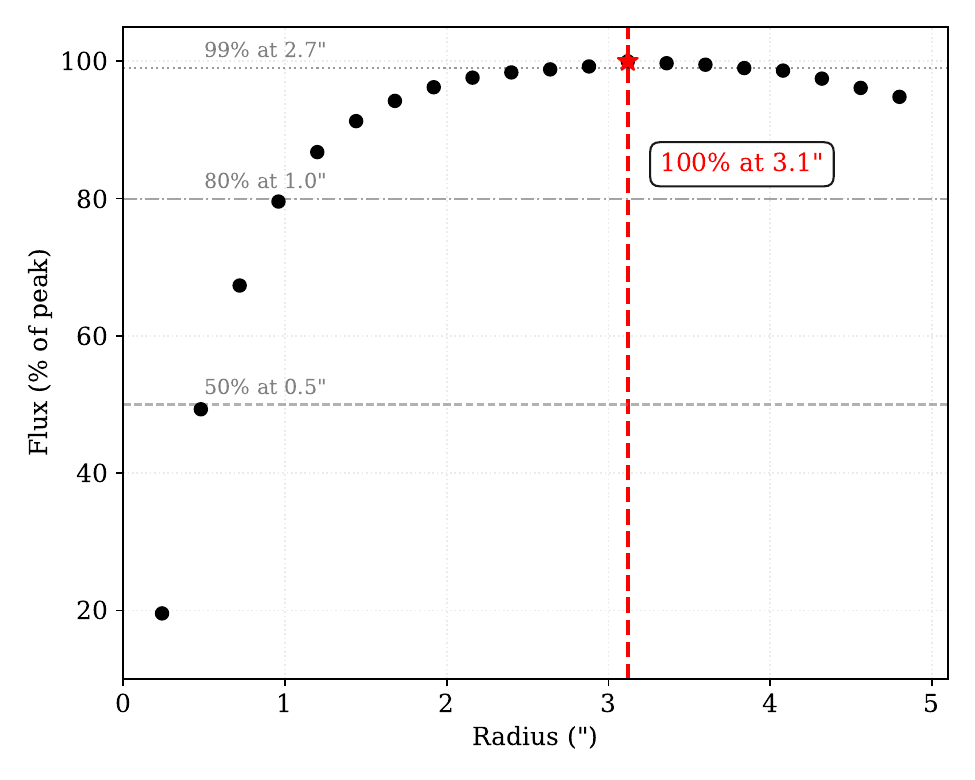}
    \put(87.5,13.5){\large \textbf{(d)}}
\end{overpic}

\vspace{0.5cm}

\caption{\textbf{(a)} Distortion pattern across the focal plane. The pixel scale is 0.485\arcsec\,pixel$^{-1}$, and the distortion amplitude remains below 0.05\%.
\textbf{(b)} Astrometric accuracy for 1084 stars in NGC~6819, showing astrometry RMS accuracy of 0.06\arcsec\ in both RA and DEC relative to Gaia~EDR3.
\textbf{(c)} Lucky-imaging PSF. A Moffat profile, with its shape parameter $\beta$ converging at $\sim1.5$, yields a FWHM of $0.5''$, providing a more robust characterization of the extended wings than a Gaussian fit. A logarithmic y-axis is used to emphasize the structure of the outer wings.
\textbf{(d)} Encircled-energy growth curve, with the EE80 radius of approximately 1\arcsec\ (about 2\,pixels).
}
\label{fig:psf}
\end{figure*}

\subsection{Astrometry and Field Distortion}\label{sec:Astrometry}

Astrometric calibration was performed using \texttt{SCAMP} with the Gaia Early Data Release~3 (Gaia~EDR3) catalog \citep{2006ASPC..351..112B,2021A&A...649A...1G}. We observed the open cluster NGC~6819 and cross-matched 1084 stars with a FoV of 10\arcmin $\times$ 10\arcmin. The measured pixel scale is 0.485\arcsec\,pixel$^{-1}$, and the geometric distortion across the focal plane is below 0.05\%, as illustrated in Figure~\ref{fig:psf}~(a).
The resulting astrometric RMS accuracy is 0.06\arcsec\ in both right ascension (RA) and declination (DEC), as shown in Figure~\ref{fig:psf}~(b). In addition, we calculated the internal astrometric RMS precision between two exposures to be 0.03\arcsec\ . Both the astrometric accuracy and precision therefore reach the sub-pixel level — approximately 12\% and 6\% of a pixel, respectively. Such sub-pixel performance is essential for the identification of transients in crowded stellar fields and facilitates high-precision cross-matching with multi-wavelength catalogs. Moreover, the high internal precision enables potential studies of stellar proper motions and precise differential astrometry for moving solar system objects within our survey fields.

\subsection{Optical Quality}\label{sec:iq}

To minimize the effect of atmospheric seeing, we employed lucky imaging~\citep{2006A&A...446..739L} to evaluate the optical quality of the system. We obtained 1500 consecutive 10\,ms exposures of Capella and selected the best 5\% based on full width at half maximum (FWHM). These frames were stacked to produce the characteristic point spread function (PSF) shown in Fig.~\ref{fig:psf}~(c). A Moffat fit to the stacked PSF yields FWHM$_{\rm meas} = 1.1$\,pixel, corresponding to $\sim 0.54\arcsec$.  
The measured PSF width is slightly overestimated due to two effects. First, the pixel size introduces undersampling, effectively convolving the PSF with a square (box) profile. This contributes a variance of $\sigma_{\rm pix}^2 = p^2/12$, where $p = 0.485\arcsec$ is the pixel scale \citep{2006hca..book.....H}. Second, the stacking procedure, which aligns images using the brightest pixel of each star as the centroid, introduces an additional alignment error of $\sigma_{\rm align} \sim 0.1\,p$ \citep{1999PASP..111..227L}. In additon, the relation between FWHM and standard deviation is ${\rm FWHM} \sim 2.355\,\sigma$ \citep{2006hca..book.....H}. The true PSF width can then be estimated by propagating these contributions in quadrature:
\[
\sigma_{\rm meas}^2 = \sigma_{\rm true}^2 + \sigma_{\rm pix}^2 + \sigma_{\rm align}^2 \,,
\]
which gives FWHM$_{\rm true} \sim 0.41\arcsec$.
For comparison, the theoretical diffraction limit is computed from the Airy disk formula, $\theta_{\rm Airy} = 1.22\,\lambda/D$, with $\lambda = 1.25~\mu$m and $D = 0.8$\,m. Accounting for this effect, the theoretical diffraction-limited FWHM is $\sim 0.41\arcsec$, in reasonable agreement with our deconvolved measurement.
The encircled-energy curve in Fig.~\ref{fig:psf}~(d) shows that the system reaches EE80 at a radius of 1\arcsec\ (approximately two pixels).

\subsection{Zero point and Overall throughput}\label{sec:efficiency}

Flux calibration was performed using several open clusters present in the Two Micron All-Sky Survey \citep[2MASS,][]{2006AJ....131.1163S}. The clusters included NGC~6633, NGC~6709, NGC~6819, NGC~6940, and NGC~7209, yielding a total of 2984 cross-matched stars. Photometry was performed by SExtractor \citep{1996A&AS..117..393B}. Using unsaturated sources with S/N $\ge$ 30, we obtained photometric zero points for flux in units of ADU (and $e^{-}$\,s$^{-1}$) of 23.27 $\pm$ 0.04\,mag (21.86\,mag) and 25.79 $\pm$ 0.03\,mag (21.78\,mag) for the INS Mars640 and YNAOIR cameras, respectively. We then used the standard flux density of 193.1\,photons\,cm$^{-2}$\,s$^{-1}$\,\AA$^{-1}$ \citep{1998A&A...333..231B} for a zero-magnitude source to estimate the theoretical photon flux of the stars. This yields overall throughputs of 36\% and 33\% for the INS Mars640 and YNAOIR systems, respectively. These efficiencies include contributions from atmospheric transmission, the filter, the optics, and the quantum efficiency of the camera. For comparison, the 4.1\,m VISTA telescope has a measured $J$ band throughput of 53\% \citep{2010SPIE.7735E..1JD}. 
Obviously, our overall efficiency is lower. This discrepancy may be attributed to the additional reflection in our Nasmyth configuration or variations in the reflective efficiency of the primary and secondary mirrors. Furthermore, 
our value is affected by several factors that introduce errors, such as the fact that the atmospheric extinction was not corrected to the zenith, and the influence of the detector's inter-pixel capacitance \citep[IPC;][]{2004SPIE.5167..204M}. In future work, we will further investigate these effects and incorporate the relevant corrections into our official data reduction pipeline.

\subsection{Photometric Depth and Noise Characterization}\label{sec:photomerty}

\begin{figure*}
	\centering
	\begin{overpic}[scale=0.45]{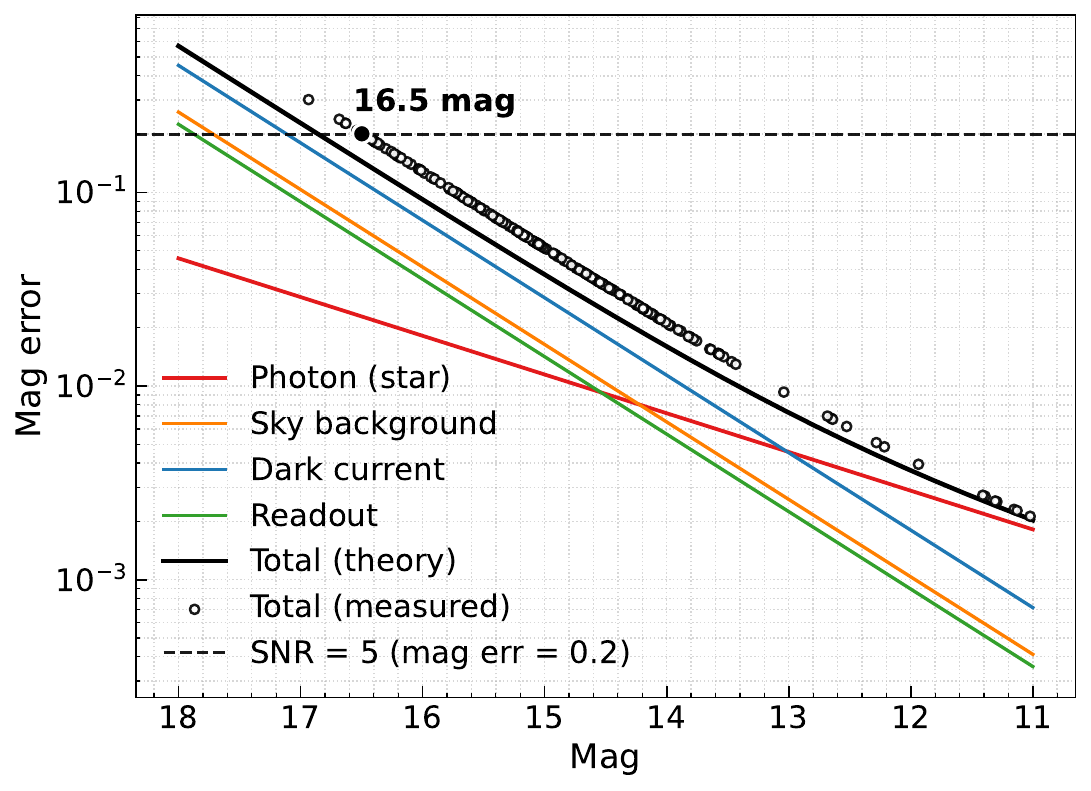}
	    \put(90,66){\large \textbf{(a)}}
	\end{overpic}
	\begin{overpic}[scale=0.45]{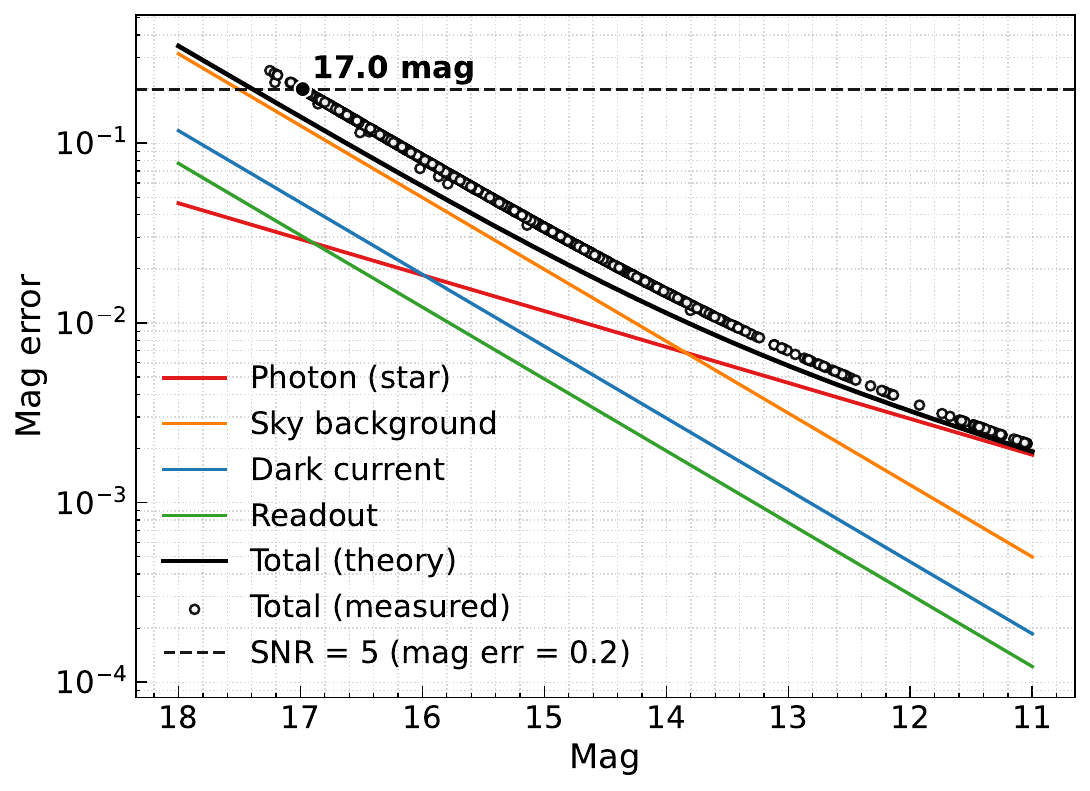}
	    \put(90,66){\large \textbf{(b)}}
	\end{overpic}
	\begin{overpic}[scale=0.2735]{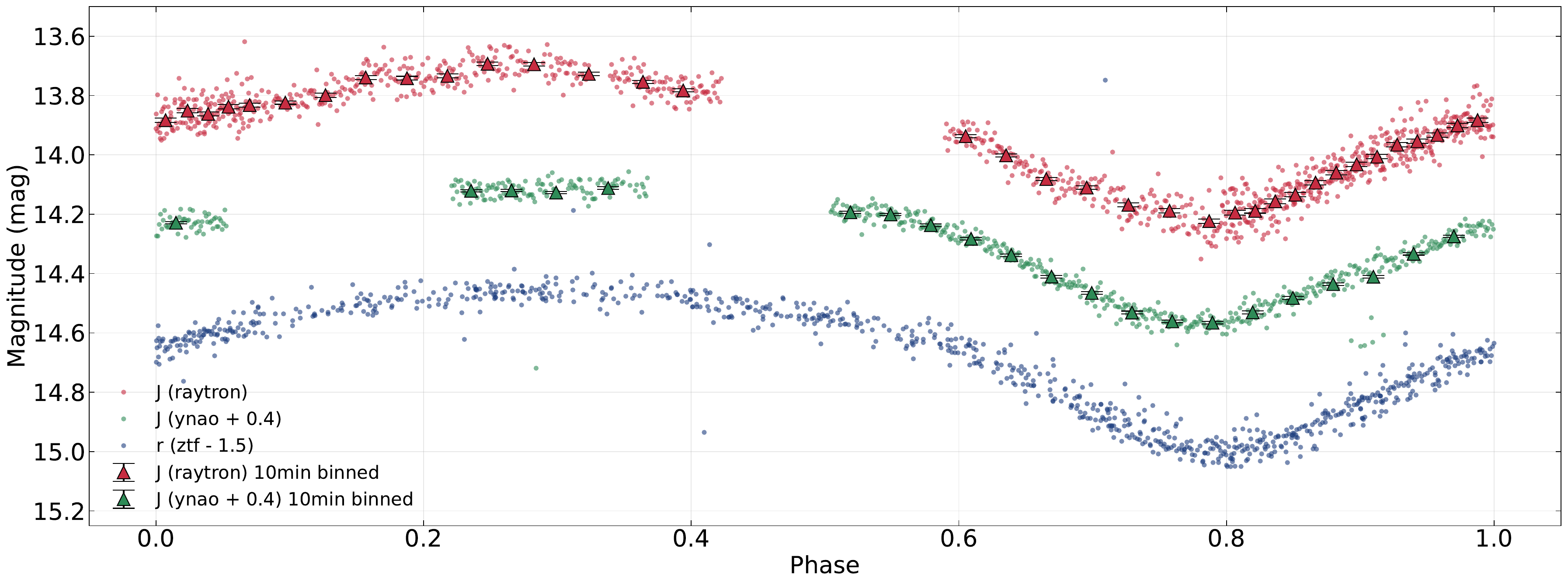}
	    \put(95.5,34.5){\large \textbf{(c)}}
	\end{overpic}
	\hspace*{-0.1cm}
	\caption{Photometric precision analysis.
	\textbf{(a)} Noise characterization of 20\,s on-sky exposures obtained with the INS Mars640 camera. The noise is dominated by dark current.
	\textbf{(b)} Noise characterization of 20\,s on-sky exposures obtained with the YNAOIR camera. The noise is dominated by sky background.
	\textbf{(c)} Light curves of CSS~J040017.9+382426 obtained with the SYSU 80\,cm infrared telescope using the INS Mars640 camera on 2024 October 15 (red) and the YNAOIR camera on 2025 September 30 (green). ZTF $r$ band history data are shown in blue. Dots denote single 20\,s exposures and triangles indicate 10\,minute integrations. For this $J=13.9$\,mag source, the INS Mars640 camera delivers 40\,mmag per frame and 7\,mmag over 10\,minutes, while YNAOIR achieves 25\,mmag per frame and 4\,mmag over 10\,minutes.}
	\label{fig:on-sky_noise}
\end{figure*}

To assess the detection depth, we analyzed the noise properties and limiting magnitude of single exposures of NGC 6819. Photometry was performed using a 1.94\arcsec\ diameter (4\,pixel) aperture, as shown in Fig. \ref{fig:on-sky_noise}.

(1) INS Mars640 camera. For sources fainter than $\sim$\,13\,mag, the total photometric noise in the middle gain mode is dominated by the dark current (221\,e$^{-}$\,s$^{-1}$\,pix$^{-1}$). For sources fainter than $\sim$\,14.5\,mag, the readout noise (33\,e$^{-}$) exceeds the photon noise, indicating that reducing the readout noise is another important objective, in addition to suppressing the dark current.

(2) YNAOIR camera. Deep cooling to 80\,K reduces the dark current to 14\,e$^{-}$\,s$^{-1}$\,pix$^{-1}$, about an order of magnitude below the sky background. Hence, future improvements in accuracy should focus on improving the stability of the system and the data reduction process, rather than on further reductions in dark current or readout noise.

For both cameras, the measured noise is $\sim$\,40\% higher than the theoretical expectation, likely owing to additional, unmodeled noise sources such as excess noise that increases with integration time \citep{2024SPIE13103E..1MD}. We will carefully evaluate and quantify these noise sources in future work. Meanwhile, our results are guiding hardware-level upgrades designed to suppress these noises at the source.

With the INS Mars640 and YNAOIR cameras, the limiting magnitudes for single 20\,s exposures are 16.5\,mag and 17.0\,mag, respectively. To reach deeper limits, we employ image stacking. With a total integration time of 30\,min, the measured detection limits reach 18\,mag for INS Mars640 and 19.4\,mag for YNAOIR. The YNAOIR depth is in good agreement with the theoretical value, but the INS Mars640 shows a gap compared to its theoretical expectation. The discrepancy between the measured results and theoretical limits likely stems from residual noise introduced during the image reduction and stacking process, such as imperfect flat-fielding, background subtraction errors, and sub-pixel registration offsets. Sub-optimal PSF screening or matching may also contribute to the reduced depth. Finally, the omission of sky brightness variability in the theoretical calculations may further account for this difference. We will optimize the data reduce for single frame and stacking procedures to improve the final depth. Furthermore, residual differences may be related to detector stability, as discussed in Section~\ref{sec:stability}.

\subsection{Photometric stability and precision}\label{sec:stability}

To assess the on-sky photometric stability, we observed CSS~J040017.9+382426, a G7-type EW eclipsing binary \citep{2014ApJS..213....9D, 2017RAA....17...87Q}. The target has a $J$ band magnitude of 13.9, a light-curve amplitude of 0.55\,mag, and an orbital period of 5.55\,hours. We obtained 7\,hours of continuous monitoring with the INS Mars640 camera on October~15, 2024, and 4.9\,hours with the YNAOIR camera on September~30, 2025, as shown in Figure~\ref{fig:on-sky_noise}~(c). Differential photometry was used to extract the light curves. The gaps in phase between 0.4 and 0.6 in both light curves are caused by the target passing through the telescope zenith avoidance region, where observations were suspended. The gap near phase~0.1 in the YNAOIR data arises from incomplete temporal coverage. A comparison with Zwicky Transient Facility (ZTF) $r$ band data \citep{2019PASP..131a8002B} reveals consistent light-curve morphologies, confirming the reliability of our photometric measurements. The light curves exhibit noise characteristics that differ from those measured in individual exposures. For the target $\sim$\,14\,mag, the INS Mars640 camera achieves a single-frame precision of 20\,mmag in Sec.~\ref{sec:photomerty}, yet the precision degrades to 40\,mmag in the continuous time-series sequence. This indicates the presence of additional noise sources during time-series observations. These excess variations may arise from unmodeled extra noise of camera, sky-background fluctuations, residual flat-field errors, and PSF variations, all of which can degrade the precision of light-curve measurements. When stacking 10\,minute integrations (20\,s $\times$ 30 frames), the precision improves to 7\,mmag — consistent with the theoretical expectation. This behavior suggests that the additional noise is still dominated by random components, whose combined noise decreases with the expected $1/\sqrt{N}$ scaling. A clearer understanding of the extra noise will require additional testing and characterization of both the detector and the readout electronics. The YNAOIR camera shows a similar trend but delivers even better photometric performance, achieving 25\,mmag per frame ($\sim$\,14\,mmag in the individual-frame test) and 4\,mmag over 10-minute integrations in the light curve.

\section{Preliminary Scientific Observations}\label{sec:scientific}

Since entering the commissioning phase in October 2024, all daily observations and operations have been conducted remotely via network connections. Initial observations were performed manually through a graphical user interface, but since January 2025 we have been progressively developing a semi-automated observing system based on the Telescope Array Operating System (TAOS)\footnote{Available at: \url{http://github.com/huyi-naoc/AAOS.git}.}\citep{10.1117/12.2313015}. We are currently working to achieve fully autonomous and embodied-intelligence operations. This includes a real-time scheduling system for the telescope; an automated response, follow-up, and feedback pipeline connected to the Einstein Probe (EP; \citealt{2015arXiv150607735Y}) and SVOM (\citealt{2016arXiv161006892W}) space--ground alert networks; automated control systems for the dome, telescope, and cameras; integrated perception and feedback from meteorological sensors and an all-sky camera; and real-time image-quality assessment, data processing, and database systems.

Our current scientific objectives are as follows:
(1) Follow-up of transients. This includes rapid target-of-opportunity (ToO) observations, long-term monitoring of supernovae and novae, and tracking of asteroids and comets from discovery to disappearance.
(2) Monitoring of variable sources. Observations of variables are typically proposed by collaborating principal investigators (PIs). We have established partnerships with more than a dozen institutions and observatories to carry out such programs. The targets include bright AGNs, high-redshift quasars, intermediate-mass black holes, as well as specific stellar objects such as brown dwarfs and RR~Lyrae stars. Additionally, we conduct repeated, long-term monitoring of selected stellar clusters and sky fields to search for previously unidentified variable sources.
(3) Static deep-field imaging for targets of interest. We will perform deep imaging of scientifically interesting regions. These observations typically involve several hours of continuous monitoring per field, reaching limiting magnitudes beyond 20.5\,mag.

\begin{figure*}
	\centering
	\begin{overpic}[scale=0.266]{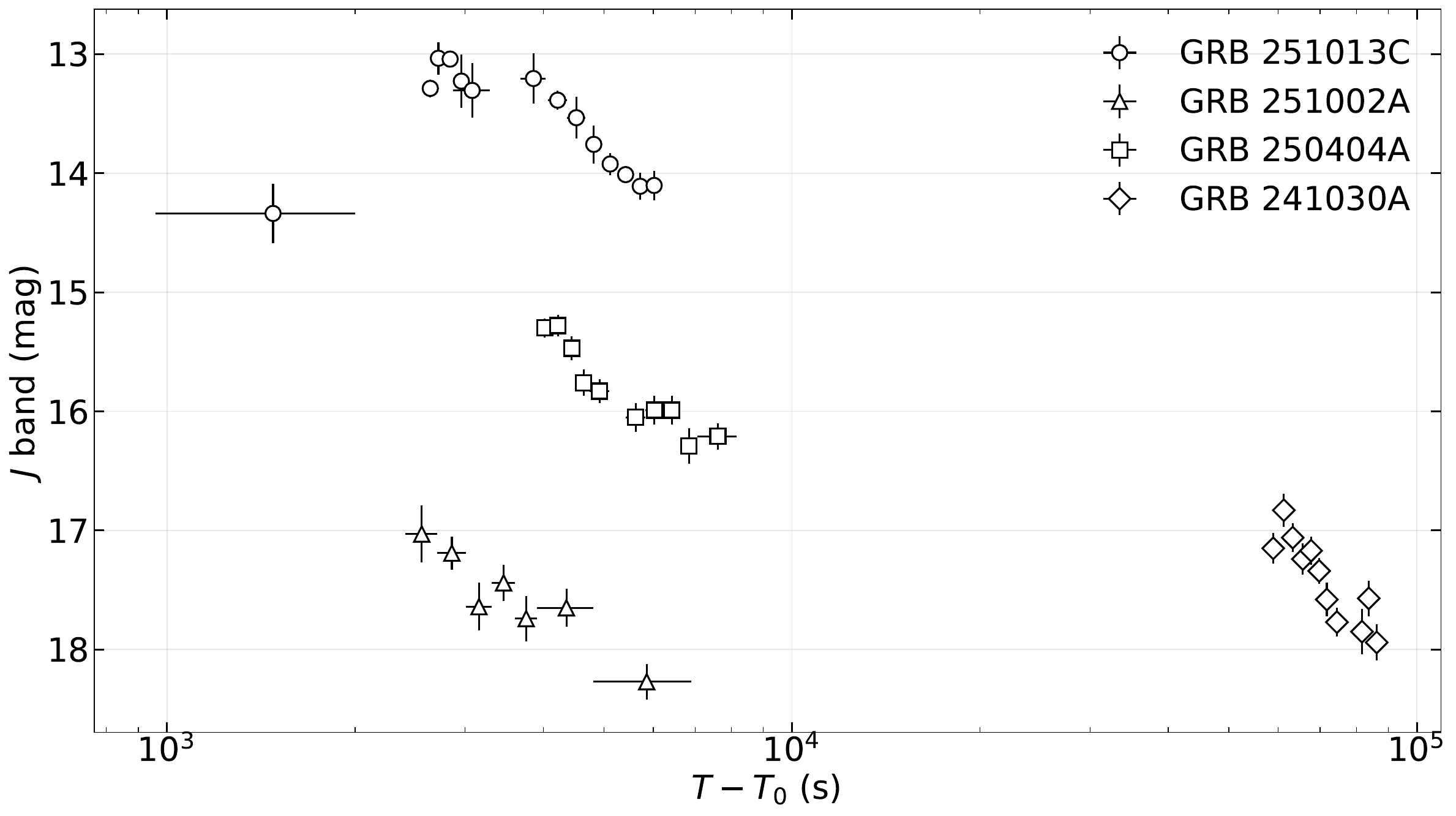}
	    \put(7,7.5){\large \textbf{(a)}}
	\end{overpic}
	\hspace*{-0.75cm}
	\begin{overpic}[scale=0.28]{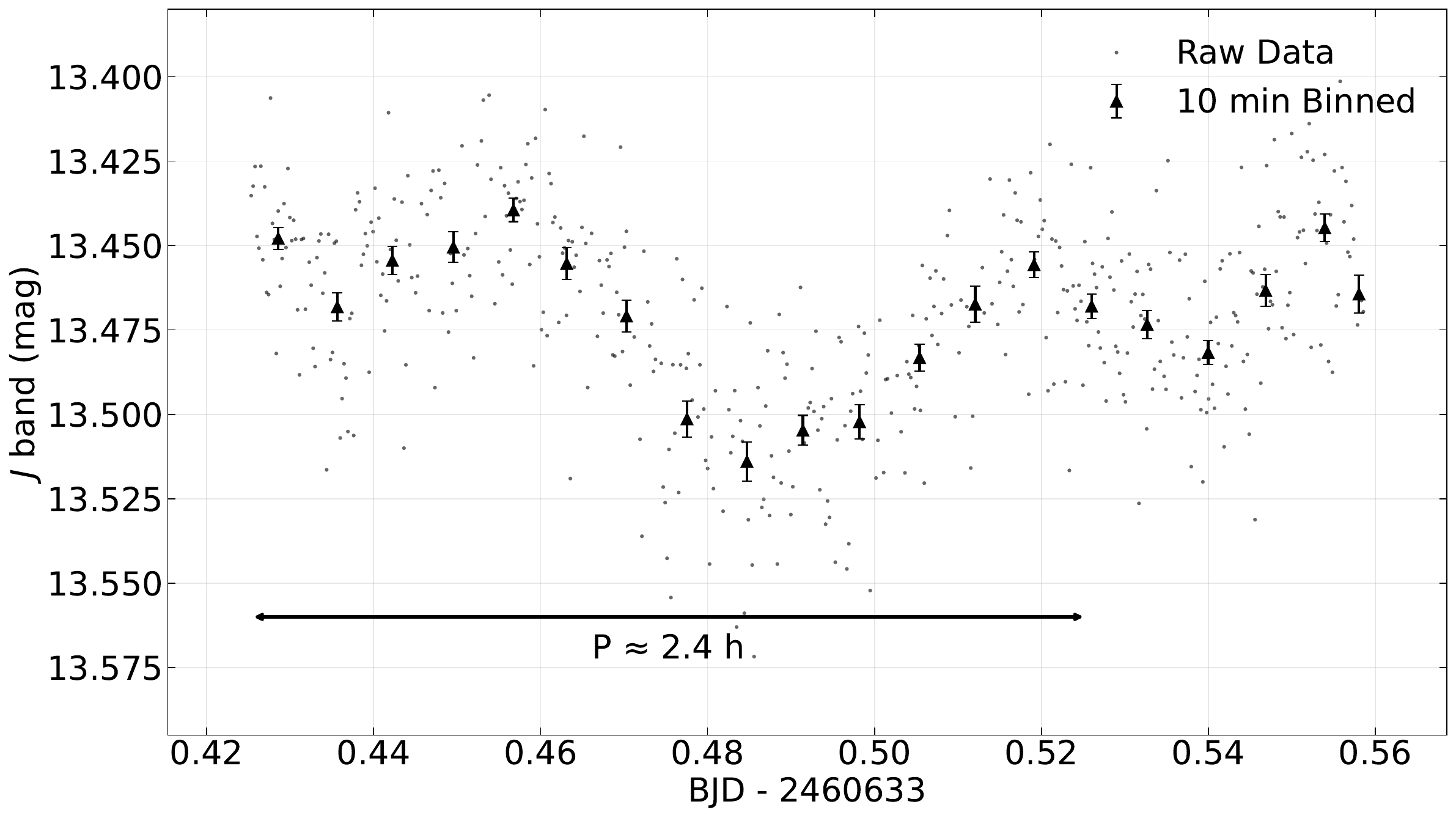}
	    \put(12.25,7.25){\large \textbf{(b)}}
	\end{overpic}
	\caption{\textbf{(a)} Light curves of GRB infrared afterglows detected by the SYSU 80\,cm infrared telescope. \textbf{(b)} Light curve of SIMP J013656.5+093347 obtained on November 18, 2024, spanning approximately 3.1\,hours.}
	\label{fig:lc}
\end{figure*}

\begin{figure*}
	\centering
	\begin{overpic}[scale=0.36]{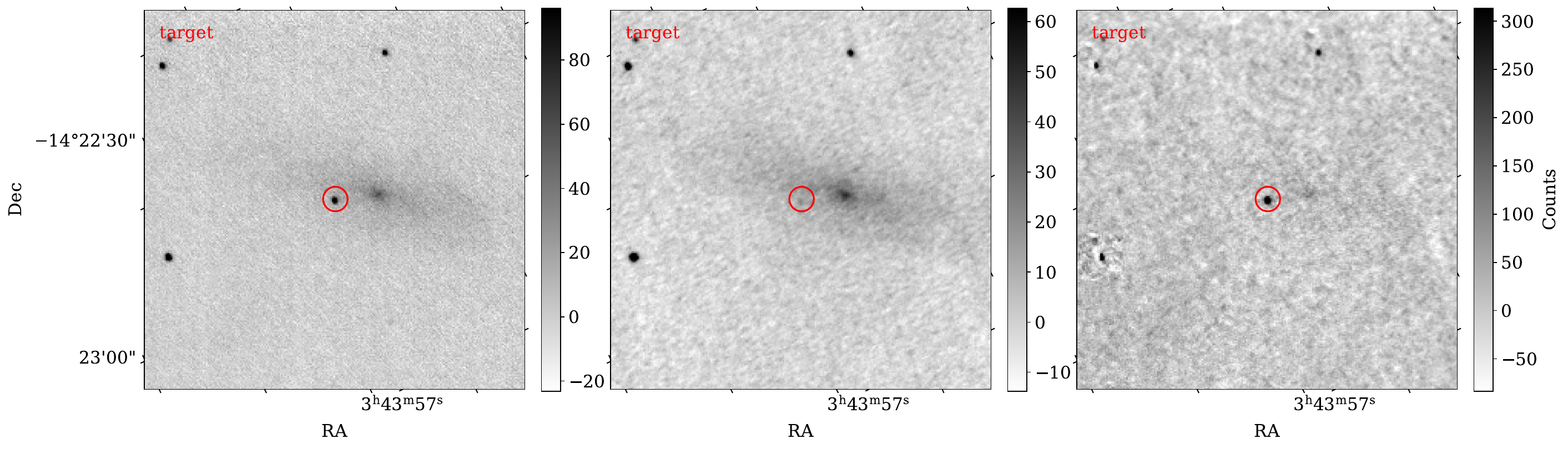}
	    \put(10,5){\large \textbf{(a)}}
	\end{overpic}
	\hspace*{0.85cm}
	\begin{overpic}[scale=0.375]{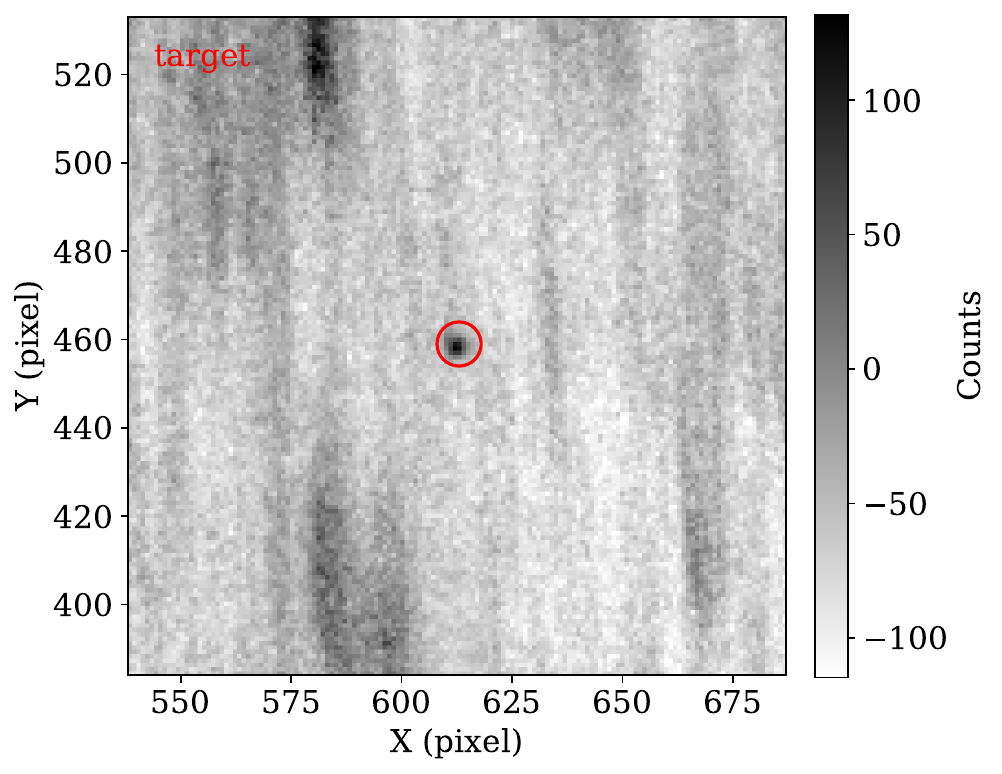}
	    \put(15,13){\large \textbf{(b)}}
	\end{overpic}\\
	\begin{overpic}[scale=0.36]{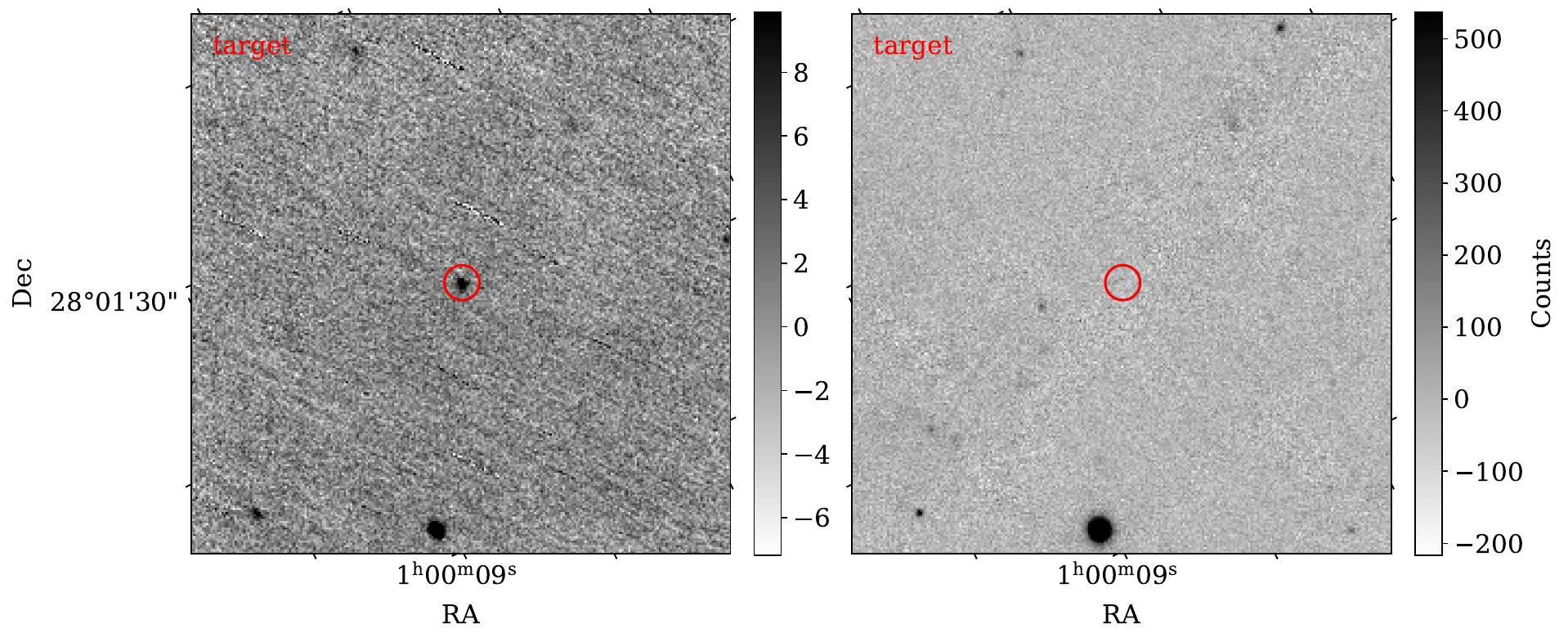}
	    \put(13,7.5){\large \textbf{(c)}}
	\end{overpic}
	\hspace*{-0.35cm}
	\begin{overpic}[scale=0.362]{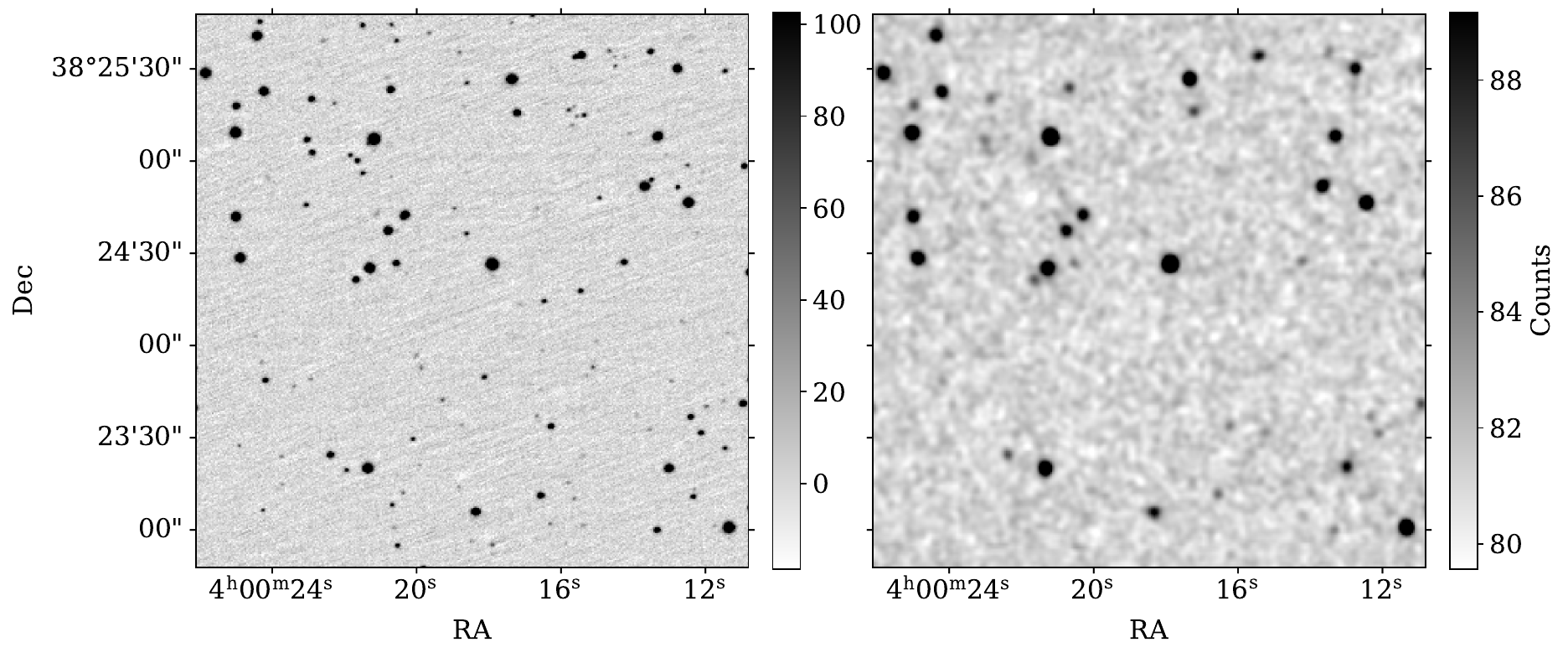}
	    \put(13,7.5){\large \textbf{(d)}}
	\end{overpic}
    \caption{SYSU 80cm $J$ band images
    \textbf{(a)} Image subtraction demonstration for SN\,2024xal. \emph{Left:} Science (sample) image obtained on 2024 October 27 with the SYSU 80\,cm telescope. \emph{Middle:} Reference (template) image taken on 2024 December 21, in which SN\,2024xal is not detected to a 5$\sigma$ limiting magnitude of 18.5\,mag. \emph{Right:} Subtracted image (science $-$ template) clearly revealing the supernova. 
	\textbf{(b)} Stacked image of 3I/ATLAS, produced by shifting 60 exposures of 20\,s to align with the comet’s position on 2025 July 14. The comet is clearly detected, while the streaks are background stars.
	\textbf{(c)} \emph{Left:} $J$ band image of the quasar J0100$-$28023 ($z=6.3$) obtained with the SYSU 80\,cm Infrared Telescope on 2024 November 1, with $J=16.3$\,mag. \emph{Right:} Archival Pan-STARRS $g$ band data, where the source is undetected down to a limiting magnitude of $\sim$\,23\,mag. High-redshift quasars invisible in the optical can be detected in the near-infrared.
	\textbf{(d)} Portion of the stacked deep image centered on RA 60.077$^\circ$, DEC 38.405$^\circ$. \emph{Left:} Stack combining individual exposures totaling 3\,hours, reaching a 5$\sigma$ depth of $J\,\sim\,20.5$\,mag. \emph{Right:} Archival 2MASS 6$\times$ Deep Survey $J$ band data, with a limiting magnitude of $\sim$\,17\,mag.
	}
	\label{fig:Observations}
\end{figure*}

\subsection{Transients}\label{sec:transient}

\subsubsection{Gamma-ray burst infrared afterglows}\label{sec:GRB}

Rapid-response optical and infrared follow-up is essential for capturing the earliest afterglow emission of Gamma-Ray Bursts. During the commissioning phase, we successfully detected infrared counterparts in four events: GRB~241030A, GRB~250404A, GRB~251002A, and GRB~251013C \citep{2024GCN.38275....1D, 2025GCN.40088....1C, 2025GCN.42078....1C, 2025GCN.42248....1Z}, as presented in Figure~\ref{fig:lc}. 
Notably, our observations of GRB~250404A contributed to the multi-wavelength analysis by \citet{2025ApJ...989L..39Y}. This study identified a distinct spectral hardening in the X-ray band, interpreting it as the signature of the external-shock afterglow onset. Our rapid $J$ band photometry provided low-energy flux constraints during this transition phase, helping to anchor the spectral energy distribution and confirm the external-shock origin of the broadband emission. Currently, our alert response still relies partially on human judgment and has not yet achieved full automation, making it difficult to reach response time on the order of tens to hundreds of seconds. For GRB~251013C, we commenced observations within 1000\,s of the alert. In particular, a NIR bump was detected, capturing the complete progression of its rising phase. In contrast, for GRB~241030A, the alert was received during the daytime, and for such events we initiate observations as soon as possible after nightfall. In addition to these detections, we responded to more than 30 alerts, most of which yielded flux upper limits of $J \sim 17.5$--$19$\,mag ($\mathrm{S/N}=5$). These non-detections nevertheless provide valuable scientific constraints. For example, our deep limits on the early-time optical--near-infrared emission of EP250702a assisted in constraining its broadband spectral properties~\citep{2026SciBu..71..538L}.

\subsubsection{Supernovae}\label{sec:SN}

Type Ia supernovae provide the most robust means of measuring extragalactic distances, exhibiting particularly high precision in the near-infrared (NIR). However, the current sample of publicly available SNe\,Ia with $J$ band light curves is limited to only $\sim$\,340, significantly fewer than those available at optical wavelengths \citep{2025arXiv251203695P}. Consequently, expanding this NIR dataset — especially for SNe\,Ia — is one of the primary scientific objectives of our program. During the commissioning phase, we monitored more than 20 transients, successfully obtaining well-sampled light curves for at least ten confirmed supernovae. Photometry of supernovae is frequently complicated by contamination from host-galaxy light. To mitigate this effect, we employ image-subtraction techniques, utilizing late-time images — obtained after the supernova has faded below the detection threshold — as background templates. A representative example is SN\,2024xal. SN\,2024xal occurred in the intermediate spiral galaxy PGC 13716 (also known as MCG-03-10-042) at a redshift of $z \sim 0.0053$. We observed this target from 2024 October 22 to December 21, securing 26 epochs down to a limiting magnitude of 18.5\,mag (S/N = 5). While the source was initially blended with the structured background of its host galaxy, it became clearly identifiable in the subtracted frames as shown in Figure~\ref{fig:Observations}~(a).

\subsubsection{Comet 3I/ATLAS}\label{sec:comet}

3I/ATLAS, the third confirmed interstellar object, was discovered on 2025 July 1 \citep{2025ApJ...989L..36S}. This approach offers a rare opportunity to directly compare interstellar material with Solar System comets and to test the universality of planet-formation processes beyond our own system. We initiated continuous monitoring shortly after the discovery. We have obtained more than 50 nights of usable data from 2025 July 3 to date, covering its ongoing observable window, with daily coverage ranging from 0.7 to 3\,hours. At the beginning, the target was faint and difficult to identify in single exposures. Using ephemerides retrieved from the NASA JPL Horizons system, we shifted and stacked our images according to the comet’s expected position. The stacked result (Figure~\ref{fig:Observations}~(b)) shows a clear detection of the comet, while the streak-like features originate from background stars elongated by the fixed-position stacking.

\subsection{Variables}\label{sec:variables}

As a general-purpose facility, the telescope also supports scientific observations of ordinary variables. In addition to our own programs, we have established collaborations with more than a dozen institutes and observatories in China and abroad. Proposed targets — each requiring different photometric depth, precision, and cadence — are reviewed and approved by the Scientific Committee of the SYSU 80\,cm Infrared Telescope. Current programs include extragalactic studies such as monitoring bright AGNs, high-redshift quasars ($z \gtrsim 6$), and intermediate-mass black holes; stellar astrophysics targets including late-type M dwarfs, RR~Lyrae stars, and maser sources; and exoplanet observations. We place particular emphasis on sources with distinct properties in the near-infrared.

\subsubsection{Bright Active Galactic Nuclei}\label{sec:agn}

Bright AGNs represent one of the key targets of our long-term monitoring program. To accurately characterize their variability and constrain the geometry of the central engine, high-cadence data are essential. We have therefore implemented a strategy of daily monitoring for selected targets. To maximize observing efficiency, we utilize the twilight window to observe bright sources ($\sim$\,10\,mag). This strategy allows us to achieve the required photometric precision without consuming prime dark time. An example of our recent monitoring is NGC 4151, a prototypical changing-look AGN that has recently undergone dramatic changes in its accretion state. While recent intensive optical reverberation mapping has successfully resolved the complex kinematics of its broad-line region (BLR) \citep{2022ApJ...936...75L, 2024ApJ...976..176F}, extending this monitoring into the near-infrared is critical for probing the reverberation signals from the hot dust torus. Of particular scientific interest is the relationship between the continuum time lag and the AGN luminosity. Recent studies indicate that the evolution of the time lag relative to luminosity is non-linear and deviates from simple theoretical predictions during distinct activity phases \citep{2025arXiv251218276F}. Continuous $J$ band monitoring is therefore vital to robustly quantify this lag-luminosity relationship, measuring the thermal dust reverberation lag and tracking the breathing of the inner dust sublimation radius during outbursts.

\subsubsection{High-redshift Quasars}\label{sec:qso}

High-redshift quasars ($z \gtrsim 6$) are vital probes of the early universe and the growth of supermassive black hole. Cosmological time dilation stretches their intrinsic variability timescales by a factor of $(1+z)$. Therefore, rather than daily monitoring, we adopt a long-cadence strategy (e.g., weekly) to efficiently track their secular evolution. A representative target of our recent observations is SDSS J0100$+$2802 at $z \sim 6.3$, a hyper-luminous quasar powered by a 12-billion-solar-mass black hole \citep{2015Natur.518..512W}. As shown in Figure~\ref{fig:Observations}~(c), this source is an optical dropout, undetectable in the Pan-STARRS $g$-band (limit $\sim$\,23.3\,mag) due to neutral hydrogen absorption, yet it is bright in the near-infrared \citep{2016arXiv161205560C}. Our $J$ band light curves will provide a reliable continuum baseline, which is essential for calibrating future multi-epoch spectroscopic observations.

\subsubsection{Brown Dwarfs}\label{sec:bd}

Brown dwarfs bridge the gap between stars and giant planets. A defining characteristic of these objects is their widespread photometric variability, which is primarily driven by the dynamic and inhomogeneous nature of their atmospheres. Their light curves frequently exhibit significant evolution over time \citep{Artigau2009, Metchev2015, 2017Sci...357..683A}. An example of our current observation is SIMP J013656.5+093347 (hereafter SIMP0136), a nearby T2 dwarf with a rapid rotation period of $\sim$\,2.4\,hours\,\citep{2006ApJ...651L..57A}. Previous studies have established it as a benchmark for variability, identifying evolving weather patterns on timescales distinct from its rotation \citep{2009ApJ...701.1534A, 2013ApJ...768..121A}. Figure~\ref{fig:lc}~(b) presents our continuous monitoring of this source over 3.1\,hours. The light curve displays a significant peak-to-peak amplitude of $\sim$\,0.07\,mag. Crucially, because our observation span exceeds one full rotation period, we effectively captured the transition between cycles. The profiles of adjacent cycles do not perfectly overlap, providing direct evidence of rapid atmospheric evolution that supports the planetary-scale wave scenario.

\subsection{Targeted Deep Imaging}\label{sec:ds}

We conduct targeted deep-field observations to supplement existing near-infrared archives. While 2MASS provides all-sky coverage, it is limited to bright targets ($J\,\le\,16$\,mag). The deeper UKIRT Hemisphere Survey (UHS; $J \sim 19.6$\,mag) offers significant improvement but lacks complete coverage of the northern sky \citep{2018MNRAS.473.5113D}. Consequently, our deep imaging program fills critical gaps, particularly in fields of high scientific interest. Such depth is indispensable for uncovering faint, red populations---such as high-redshift quasars and ultracool dwarfs---that fall below the detection thresholds of shallow wide-field surveys \citep[e.g.,][]{2016ApJ...828...26M}. Furthermore, it enables the study of stellar mass assembly in dusty star-forming galaxies at ``Cosmic Noon'' ($z \sim 1$--3) \citep[e.g.,][]{2014ARA&A..52..415M} and serves as essential reference templates for identifying elusive, dust-enshrouded transients in difference imaging analyses \citep[e.g.,][]{2019ApJ...886...40J}. Deep images are constructed through both dedicated long-exposure observations of specific targets and the cumulative co-addition of multi-epoch time-domain monitoring data. As demonstrated in Figure~\ref{fig:Observations}~(d), by selecting high-quality frames (FWHM $< 4$\,pixels) from a total of 3.9\,hours of monitoring data, we generated a 3-hour stack reaching a depth of $J \sim 20.5$\,mag. This depth is comparable to the NOAO Deep Wide-Field Survey (NDWFS; \citealt{2013ApJ...772...26A}), which reaches $J \sim 20.2$\,mag. With our current field of view, dedicating $\sim$\,20\% of annual observing time to such cumulative integration would yield approximately 2\,deg$^2$ of deep coverage per year.

\section{Summary}\label{sec:summary}

The Sun Yat-sen University (SYSU) 80\,cm near-infrared telescope was commissioned in October 2024 at the Lenghu site and has since commenced routine scientific operations. One of the primary objectives of the facility is the rapid follow-up of transients---including gamma-ray bursts (GRBs) and supernovae---and the long-term monitoring of variable sources such as AGNs and cool dwarfs. A key operational feature is the telescope’s high-speed slewing capability, which enables target acquisition within $\sim$\,15\,s across the sky. This agility, combined with a blind-pointing accuracy better than 10\arcsec, makes the system uniquely suited for capturing the earliest phases of fast-evolving transients on sub-minute timescales. 

The initial operations utilized a commercial-grade InGaAs camera (INS Mars640), which was successfully upgraded in mid-2025 to a deeply cooled, science-grade camera (YNAOIR). This upgrade significantly suppressed dark current and readout noise to an order of magnitude below the near-infrared sky background. With the new system, we achieve limiting magnitudes of $J \sim 17$\,mag in single 20\,s exposures and deepen to $J \sim 19$--20\,mag via image stacking. For targets with $J \sim 14$\,mag, the system delivers photometric precision at the few-millimagnitude level, demonstrating its efficacy for high-precision time-domain studies. We have already carried out long-term monitoring of variable sources such as bright AGNs, high-redshift quasars, and brown dwarfs.

Looking ahead, we aim to further optimize the system's sensitivity and efficiency. Future work will focus on characterizing and mitigating residual instrumental noise sources, refining data reduction pipelines to improve background subtraction, and testing new generation large-format detectors. Operationally, we plan to transition from remote-assisted observing to fully robotic operations, incorporating fully automated alert parsing and decision-making to reduce response latencies within tens of seconds.

The telescope represents a new generation near-infrared (NIR) facility in China, designed to address the growing demands of time-domain astronomy. On one hand, scientific feedback from this facility directly informs the ongoing development and optimization of new InGaAs camera systems. On the other hand, it has encouraged other institutions in China to develop dedicated infrared telescopes or integrate infrared cameras into existing optical platforms. For instance, the validation of the INS Mars640 camera on our telescope paved the way for its deployment in the JinShan project \citep{2025GCN.42990....1F}. On the international stage, the facility provides a practical reference as one of the few operational InGaAs-based NIR telescopes, while its location in East Asia complements the longitudinal coverage of the global time-domain monitoring network.

\begin{acknowledgments}

We are grateful to the anonymous reviewer for the professional and insightful suggestions, which have significantly enhanced the clarity and rigor of this manuscript. The SYSU 80cm infrared telescope is operated and managed by the Department of Astronomy, Sun Yat-sen University. YH is supported by the National Key Research and Development Program of China (Grant No. 2023YFA1608301), the National Natural Science Foundation of China (Grant Nos. 12373092, 11873010, and 11733007), and the Strategic Priority Research Program of the Chinese Academy of Sciences (Grant No. XDB0550101).

\end{acknowledgments}

\begin{contribution}
\end{contribution}

\bibliography{PASPsample701}{}
\bibliographystyle{aasjournalv7}

\end{document}